\documentclass[12pt]{article}

\usepackage{textcomp}
\usepackage{chetnew}
\usepackage{amssymb,amsmath,amsfonts,mathrsfs}
\usepackage[utf8]{inputenc}
\usepackage{xcolor,enumerate}

\usepackage{bm}
\usepackage[colorinlistoftodos,textsize=tiny]{todonotes}

\def\one{{\,\hbox{1\kern-.8mm l}}}

\newcommand{\U}{\mathrm U}
\newcommand{\Tr}{\mathrm{Tr}}

\newcommand{\QQ}{\mathcal Q}

\allowdisplaybreaks

\newcommand{\dd}{\mathrm{d}}

\def\makeatletter{\catcode`\@=11}
\makeatletter
\def\mathbox#1{\hbox{$\m@th#1$}}%
\def\math@ccstyles#1#2#3#4#5#6#7{{\leavevmode
      \setbox0\mathbox{#6#7}%
      \setbox2\mathbox{#4#5}%
      \dimen@ #3%
      \baselineskip\z@\lineskiplimit#1\lineskip\z@
      \vbox{\ialign{##\crcr
             \hfil \kern #2\box2 \hfil\crcr
             \noalign{\kern\dimen@}%
             \hfil\box0\hfil\crcr}}}}
\def\mathaccstyles{\math@ccstyles\maxdimen}
\def\maththroughstyles{\math@ccstyles{-\maxdimen}}
\def\unity%
 {\maththroughstyles{.45\ht0}\z@\displaystyle {\mathchar"006C}\displaystyle 1}

\def\FF{{\cal F}}

\def\HH{{\cal H}}
\def\II{{\cal I}}

\def\LL{{\cal L}}

\def\NN{{\cal N}}

\def\QQ{{\cal Q}}

\def\ZZ{{\cal Z}}

\def\Tr{{\rm {Tr}}}

\def\beq{\begin{equation}}
\def\eeq{\end{equation}}
\newcommand{\bea}{\begin{eqnarray}}
\newcommand{\eea}{\end{eqnarray}}
\def\bal{\begin{align}}
\def\eal{\end{align}}

\preprint{QMUL-PH-19-17}

\title{AdS$_7$ Black-Hole Entropy and 5D $\mathcal{N}=2$ Yang-Mills }

\author{G.~K\'antor,$^{a,\heartsuit}$ C.~Papageorgakis\;$^{a, \spadesuit}$ and P.~Richmond\;$^{b,\diamondsuit}$}

\affiliation{$^a$CRST and School of Physics and Astronomy\\ Queen Mary
  University of London, London E1 4NS, UK\\ $ $ \\ $^b$Department of
  Mathematics\\ King's College London, London WC2R 2LS, UK

\emails{$^\heartsuit$g.kantor@qmul.ac.uk, $^\spadesuit$c.papageorgakis@qmul.ac.uk, $^\diamondsuit$paul.richmond@kcl.ac.uk}
}

\abstract{We generalise the work of 1810.11442 for the case of AdS$_7$/CFT$_6$. Starting from the 2-equivalent charge, 3-equivalent rotation non-extremal black-hole solution in 7D gauged supergravity, we consider the supersymmetric and then the extremal limit and evaluate the associated thermodynamic quantities. Away from extremality, the black-hole solution becomes complex. The entropy is then given by the Legendre transform of the on-shell action with respect to two complex chemical potentials subject to a constraint. At the conformal boundary  we derive the dual background and evaluate the corresponding partition function for the $A_{N-1}$ 6D (2,0) theory at large $N$ in a Cardy-like limit. This is carried out via a 5D $\NN=2$ super Yang--Mills calculation on $S^5$. The gravitational on-shell action is found to be exactly reproduced by the boundary partition function at large $N$. We argue that this agreement puts strong constraints on the form of possible higher-derivative corrections to the 5D gauge theory that is used in the $S^5$ evaluation. }

\date{\today}

\begin{document}

\maketitle
\toc

\hypersetup{pageanchor=true}

\setcounter{tocdepth}{2}

\section{Introduction and Summary}\label{intro}

Since the introduction of the superconformal index \cite{Kinney:2005ej,Romelsberger:2005eg,Bhattacharya:2008zy} there has been a puzzle pertaining to why it seems to only be capturing supergraviton-state contributions in AdS, and not those of certain BPS black holes as predicted by the AdS/CFT correspondence at large $N$ \cite{Grant:2008sk,Chang:2013fba}. Over the years there have been several approaches to this problem. For example, one of the early explanations appealed to the possibility of huge cancellations between bosonic and fermionic states. Later on, the reformulation of the superconformal index in $D$ even spacetime dimensions through the operator-state map as an R-twisted supersymmetric partition function on $S^{D-1}\times S^1_\beta$ revealed that \begin{equation}
  \label{eq:15}
  \ZZ_{S^{D-1}\times S^1_\beta} \ = \ e^{- \FF} \II\;,
\end{equation}
where $\II$ is the superconformal index, $\beta$ the radius of the thermal circle and $\FF$ a quantity related to the vacuum Casimir energy \cite{Nawata:2011un,Kim:2012ava,Lorenzen:2014pna,Assel:2015nca}. Unlike the index, this ``supersymmetric Casimir energy'' was shown to exhibit the expected scaling of degrees of freedom at large $N$. However matching the precise coefficient predicted by AdS/CFT, and corresponding entropy of BPS black holes in the gravity dual, turns out to be subtler, see for example \cite{Cassani:2014zwa}. More recently, the black-hole entropy for even $D$ has been recovered from field theory through the extremisation of a quantity closely resembling the supersymmetric Casimir energy \cite{Hosseini:2017mds,Hosseini:2018dob}.

Over the last few months, there has been a significant acceleration of
activity in this direction. In \cite{Cabo-Bizet:2018ehj}, a
complete gravitational derivation of the field-theoretic entropy function was performed for AdS$_5$/CFT$_4$,\footnote{For generalisations see \cite{Cassani:2019mms,Larsen:2019oll}.} while in \cite{Choi:2018hmj,Choi:2018fdc} the entropy function was reproduced from field theory in the Cardy limit of large charges, for a variety of bulk dimensions. In yet another line of attack, the superconformal index itself (and not the supersymmetric Casimir energy) was shown to be capturing larger than previously thought degeneracies for a particular complexification of chemical potentials, through a formulation that involves solutions for auxiliary Thermodynamic Bethe Ansatz (TBA) equations \cite{Benini:2018mlo,Benini:2018ywd}. Finally, the general behaviour seen in \cite{Benini:2018mlo,Benini:2018ywd} was reproduced for 4D $\NN=1$ SCFTs in the Cardy limit without resorting to the TBA equations in \cite{Kim:2019yrz,Cabo-Bizet:2019osg}.

In this note, we will focus on the relationship between the AdS$_7$ black-hole entropy and the superconformal index of the six-dimensional (2,0) theory.\footnote{Note that the AdS$_7$ entropy function was first written down in \cite{Hosseini:2018dob}, while it was reproduced in the Cardy limit in \cite{Choi:2018hmj}. However, our scope here will be to provide a microscopic derivation of this quantity from the six-dimensional CFT dual theory.} We will first generalise the AdS$_5$ analysis of \cite{Cabo-Bizet:2018ehj} for the case of 2-equivalent charge and 3-equivalent rotation non-extremal black-hole solutions. A study of the regularity conditions for the metric and Killing spinors in the bulk leads to a specific background at the boundary of AdS$_7$. The AdS/CFT correspondence then dictates that the black-hole entropy should be related to an R-twisted, supersymmetric partition function for the six-dimensional $A_{N-1}$ (2,0) theory on this particular background in the large-$N$ limit.

As the interacting (2,0) theory does not admit a Lagrangian description one cannot directly employ the method of supersymmetric localisation to evaluate the boundary partition function. We therefore turn to the existing literature, where it has instead been conjectured to arise from a 5D supersymmetric partition function on $S^5$ for the maximally-supersymmetric ($\NN = 2)$ Yang--Mills theory (MSYM) with $SU(N)$ gauge group \cite{Lockhart:2012vp,Kallen:2012cs,Kallen:2012va,Kallen:2012zn,Kim:2012ava,Kim:2012qf,Kim:2013nva,Bobev:2015kza}---the circle reduction of the (2,0) theory on $S^5 \times S_\beta^1$. A modification of these results at large $N$ and in a Cardy-like limit reproduces a (generalised) supersymmetric Casimir energy that exactly matches the gravitational prediction.\footnote{It would be very interesting to revisit the original work of \cite{Bhattacharya:2008zy} and investigate the scaling of degrees of freedom directly in the 6D superconformal index along the lines of \cite{Benini:2018mlo,Benini:2018ywd}.}

Lastly, we argue that the 6D Casimir energy is sensitive to the choice of 5D theory for which one evaluates the partition function on $S^5$. Since 5D MSYM is conventionally thought of as a Wilsonian effective field theory, it is expected to differ from the microscopic (2,0) theory in the UV by an infinite tower of higher-derivative corrections. The latter can be organised into D-terms and F-terms, out of which only the F-terms affect the $S^5$ partition-function computation \cite{Jafferis:2012iv,Chang:2014kma,Lin:2015zea,Mezei:2018url}. The precision agreement between AdS$_7$ and CFT$_6$ hence constrains F-term-correction coefficients. Setting all these coefficients to zero, as done in \cite{Lockhart:2012vp,Kallen:2012cs,Kallen:2012va,Kallen:2012zn,Kim:2012ava,Kim:2012qf,Kim:2013nva,Bobev:2015kza}, provides an obvious solution. However, since the inclusion of higher-dimension operators lead to different matrix models, we cannot prove that there is no infinite sequence of F-term corrections with non-zero coefficients that also reproduces the result for the Casimir energy. At the very least, the AdS$_7$ black-hole entropy severely restricts the form of the completion of 5D MSYM towards the 6D (2,0) theory in the UV. We also comment on the connection between our results and the conjecture that 5D MSYM, without the need for higher-derivative corrections, captures all the information about the (2,0) theory on $S^1$ \cite{Douglas:2010iu,Lambert:2010iw,Papageorgakis:2014dma}.

We should emphasise that the bulk approach of \cite{Cabo-Bizet:2018ehj} determining the black-hole entropy that we follow in this paper is an interesting recent alternative to Sen's entropy-function formalism \cite{Sen:2007qy,Sen:2008yk}. The latter employs the attractor mechanism to identify the leading degrees of freedom of the full spacetime with the Bekenstein-Hawking entropy at the black-hole horizon. Therefore, the associated extremisation principle derived in \cite{Sen:2007qy,Sen:2008yk} is expected to agree to leading order with the one used here. However, due to considering the near-horizon geometry, Sen's approach does not explicitly determine the dual supersymmetric partition function at the boundary, which is where we would like to focus our attention in this work. 

Note also that the statistical entropy of various asymptotically-AdS black holes (not only for $D$ even) can be reproduced microscopically from a different, ``topologically-twisted index''.\footnote{In this work, whenever we refer to the ``index'' we will mean the superconformal index and not its topologically-twisted version.} The latter can be evaluated through supersymmetric localisation for a topologically-twisted gauge theory along the lines of \cite{Benini:2015noa}; see e.g. \cite{Benini:2015eyy,Hosseini:2018uzp} for applications.

The rest of this note is organised as follows: In Sec.~\ref{bulk}, we analyse the 2-equivalent charge, 3-equivalent rotation black-hole solution in AdS$_7$ and determine how it fixes the form of the boundary partition function that is AdS/CFT dual to the corresponding on-shell action at large $N$. We then discuss the evaluation of this boundary partition function from 5D MSYM using supersymmetric localisation in Sec.~\ref{boundary}. Finally, in Sec.~\ref{implications} we discuss the implications of this precision matching for the UV completion of 5D MSYM by examining the contributions of higher-derivative corrections. 

\vskip .5cm

{\bf Note added:} Shortly before v1 of this manuscript appeared on the
arXiv, we received \cite{Cassani:2019mms}, which overlaps with the results of Secs.~\ref{sec:21}-\ref{sec:26}.

\section{2-equivalent-charge, 3-equivalent-rotation Black Holes in AdS$_7$}\label{bulk}

We begin with the study of supersymmetric black-hole solutions of 7D gauged supergravity (SUGRA). We will follow the AdS$_5$ analysis \cite{Cabo-Bizet:2018ehj} very closely throughout and discuss two distinct limits, that of supersymmetry and extremality. Generalising the arguments of \cite{Cabo-Bizet:2018ehj} is conceptually straightforward and we do not encounter any surprises, although the details are more involved. This fact will force us to consider black-hole metrics with 2-equivalent charges and 3-equivalent rotations.

\subsection{Non-extremal AdS$_7$ Black Holes}\label{sec:21}

The low energy limit of M-theory, 11D SUGRA, admits solutions where the geometry is of the form $\mathcal{M}_7 \times S^4$ with the manifold $\mathcal{M}_7$ being asymptotically AdS$_7$. There is a consistent truncation of 11D SUGRA on $S^4$ such that $\mathcal{M}_7$ is a solution to $\mathcal{N}=4$ $SO(5)$ gauged SUGRA in seven dimensions \cite{Townsend:1983kk}. Amongst the possible solutions there is an expectation of finding seven-dimensional black holes with two independent parameters $(\delta_1,\delta_2)$ for the charges of the $\U(1)^2$ Cartan subgroup of $SO(5)$ and three independent parameters $(a_1,a_2,a_3)$ for the rotations supported by the $SO(2)^3$ in the maximal compact subgroup of $SO(2,6)$. 

For practical purposes charged solutions are sought within a $U(1)^2$ truncation of the maximal $SO(5)$ theory. The bosonic field content of this truncation consists of the metric, two 1-form gauge potentials $A^1_{(1)},A^2_{(1)}$, one 3-form $A_{(3)}$ which may be traded for a 2-form by utilising an odd-dimensional self-duality relation and two scalars $X_1,X_2$. The most general black hole solution with generic charges and rotations has not yet been found but solutions with two charges and three equivalent rotations \cite{Chong:2004dy} or two equivalent charges and three rotations \cite{Chow:2007ts} are known.

\subsection{2-equivalent-charge, 3-equivalent-rotation Black Hole}

For simplicity, we choose to focus on a subclass within the non-extremal AdS$_7$ black holes where all the charges are set to be equal to each other, $\delta_1 = \delta_2 = \delta$, as are all the rotational parameters, $a_1 = a_2 = a_3 = a$. As a result, this family of non-extremal black holes depends on three parameters $(m,\delta,a)$. In this scenario the solutions of \cite{Chong:2004dy} and \cite{Chow:2007ts} must coincide. The solution, incorporating a correction to the original literature \cite{Chong:2004dy}, is
\begin{align}
	\dd s^2_7 \ =& \ H^{\frac{2}{5}} \bigg[ - \frac{Y}{f_1 \Xi_-^2} \dd t^2 + \frac{\Xi \rho^6}{Y} \dd r^2 + \frac{f_1}{H^2 \Xi^2 \rho^4} \left( \sigma - \frac{2 f_2}{f_1} \dd t \right)^2 + \frac{\rho^2}{\Xi} \dd s^2_{\mathbb{CP}^2} \bigg] \label{BHmetric} \, ,\\
	A^1_{(1)} \ =& \ A^2_{(1)} \ = \ \frac{2m s c}{\rho^4 \Xi H} \big( \Xi_- \dd t - a \sigma \big) + \frac{\alpha}{\Xi_-} \dd t \label{BHgauge} \, , \\
	A_{(3)} \ =& \ \frac{2m a s^2}{\rho^2 \Xi \Xi_-} \sigma \wedge ( 2 \dd \sigma ) \label{BH3form} \, ,\\
	X_1 \ =& \ X_2 \ = \ H^{-1/5} \, . \label{BHscalar}
\end{align}
As in \cite{Cabo-Bizet:2018ehj}, we have added a pure gauge term, $\Xi_-^{-1}\alpha \dd t$, to each of the $U(1)$ gauge fields. The quantities $f_1,f_2,H,Y$ appearing above are all functions of the radial coordinate $r$ through the definition $\rho^2 = \Xi r^2 $. They are given by
\begin{align}
	H \ =& \ 1 + \frac{2m s^2}{\rho^4} \, ,\\
	f_1 \ =& \ \Xi \rho^6 H^2 - \frac{(2 \Xi_+ m a s^2)^2}{\rho^4} + 2 m a^2 \big[ \Xi_+^2 + c^2 ( 1 - \Xi_+^2) \big] \label{metricOrig} \, ,\\
	f_2 \ =& \ - \frac{g \Xi_+ \rho^6 H^2}{2} + m a c^2 \, ,\\
	Y \ =& \  g^2 \rho^8 H^2 + \Xi \rho^6 - 2 m \rho^2 \big[ a^2 g^2 c^2 + \Xi \big] + 2 m a^2 \big[ \Xi_+^2 + c^2 ( 1 - \Xi_+^2) \big] \, ,
\end{align}
with $g$ the gauge coupling parameter and 
\begin{align}
	\Xi_{\pm} \ =& \ 1 \pm a g \, , \qquad \Xi \ = \ 1 - a^2 g^2 \ = \ \Xi_- \Xi_+ \, , \qquad s \ = \ \sinh \delta \, , \qquad c \ = \ \cosh \delta \, .
\end{align}
The black-hole outer horizon is located at the largest positive root of $Y(r)$ which we denote by $r_+$. The remaining data of the solution are given by
\begin{align}
	\sigma \ =& \ \dd \psi + \frac{1}{2} \sin^2 \xi l_3 \, , \quad \dd s^2_{\mathbb{CP}^2} \ = \ \dd \xi^2 + \frac{1}{4} \sin^2 \xi ( l_1^2 + l_2^2 ) + \frac{1}{4} \sin^2 \xi \cos^2 \xi l_3^2 \, , 
\end{align}
with $(l_1,l_2,l_3)$ a choice of left-invariant 1-forms for $SU(2)$. 

The solution given by \eqref{BHmetric}-\eqref{BHscalar}, along with its associated Killing vector at the horizon allow for the computation of thermodynamical aspects of
the black hole such as its temperature. This is done with respect to
the non-rotating Killing vector
\begin{align}
  \label{eq:16}
\ell \ = \  \frac{\partial}{\partial t} - g \Omega \frac{\partial }{\partial \psi} \, ,
\end{align}
found by redefining $t \rightarrow \Xi_- t$, $\psi \rightarrow \psi -g t$. In these new coordinates the temperature,\footnote{We have incorporated the corrections to \cite{Cvetic:2005zi} as noted in \cite{Hosseini:2018dob} and \cite{Choi:2018hmj}.} entropy,\footnote{Here we have corrected the expression given in \cite{Choi:2018hmj}.} angular velocity and electrostatic potential are \cite{Cvetic:2005zi} 
\begin{align}
	T \ =& \ \frac{\partial_r Y}{4\pi g \rho^3 \sqrt{\Xi f_1}} \label{Temp} \, , \\
	S \ =& \ \frac{1}{G_N} \frac{1}{4} \frac{\pi^3 \rho^2 \sqrt{f_1}}{\Xi^3} \label{Ent} \, , \\
	\Omega \ =& \ - \frac{1}{g} \left( g + \frac{2f_2 \Xi_-}{f_1} \right) \label{Omega} \, , \\
	\Phi \ =& \ \frac{4m s c}{\rho^4 \Xi H} \left( \Xi_- - a \frac{2f_2 \Xi_-}{f_1} \right) \label{Phi} \, ,
\end{align}
and are all evaluated at the outer horizon $r = r_+$. $G_N$ denotes the seven-dimensional Newton constant. 
The conserved charges, namely the energy, angular momentum and electric charge, are:
\begin{align}
	E \ =& \ \frac{1}{G_N} \frac{1}{g} \frac{m \pi^2}{32 \Xi^4} \bigg[ 12 (a g+1)^2 (a g (a g+2)-1)-4 c^2 \left(a^2 g^2 (3 a g (a g+4)+11)-8\right) \bigg] \label{En} \, , \\
	J \ =& \ - \frac{1}{G_N} \frac{m a \pi^2}{16 \Xi^4} \bigg[ 4 a g (a g+1)^2-4 c^2 \left(a^3 g^3+2 a^2 g^2+a g-1\right) \bigg] \label{J} \, , \\
	Q \ =& \ \frac{1}{G_N} \frac{1}{2g} \frac{m s c \pi^2}{2 \Xi^3} \, , \label{Q}
\end{align}
and are found by integrating the first law:
\begin{align}
	\dd E \ = \ T \dd S + 3 \Omega \dd J + 2 \Phi \dd Q  \, . \label{1stLaw}
\end{align}
We will also need the free energy, $I$, of the black hole solution. The so-called quantum statistical relation gives this as
\begin{align}
	I \ =& \ \beta( E - TS - 3 \Omega J - 2 \Phi Q ) \, , \label{QSR}
\end{align}
where $\beta=T^{-1}$ and it has been evaluated in \cite{Chen:2005zj} to be
\begin{align}
	I \ = \ \frac{\beta\pi^2}{8 \Xi^3} \bigg[ m &- g^2 r_+^6 - g^2 m s^2 ( 4 r_+^2 - a^2 ) \nonumber \\
	&- \frac{4g (m s^2)^2[g r_+^4+a^2 g(1+ag)r_+^2+2g m s^2-a^3(1+ag)^2]}{r_+^6+2 m s^2 r_+^2-2a^2(1+ag)} \bigg] \, .
\end{align}
The free energy is expected to coincide with the on-shell supergravity action evaluated on the black-hole solution.\footnote{It would be interesting to derive this explicitly in supergravity without resorting to the quantum statistical relation. } The black-hole entropy is given in terms of the Legendre transform of the on-shell action with respect to the chemical potentials $\beta, \Omega, \Phi$, conjugate to the charges $E, J, Q$ respectively:
\begin{align}
  \label{eq:17}
  E \ = \ \frac{\partial I}{\partial \beta}\;, \qquad J \ = \ -\frac{1}{3\beta}\frac{\partial I}{\partial \Omega}\;, \qquad Q \ = \ -\frac{1}{2\beta}\frac{\partial I}{\partial \Phi}\;,
\end{align}
hence
\begin{align}
  \label{eq:18}
  S \ = \ - I + \beta E  - 3 \beta \Omega J - 2 \beta \Phi Q\;.
\end{align}

\subsection{Supersymmetry}

The non-extremal black hole solution detailed in the previous section is supersymmetric whenever the charge and rotation parameter satisfy one of the following two relations
\begin{align}
	e^{2\delta} \ =& \ 1 - \frac{2}{3a g} \, , \label{SUSY} \\
	e^{2\delta} \ =& \ 1 - \frac{2}{a g} \, .
\end{align}
However, when the second condition holds, it is not possible to cloak closed timelike curves (CTCs) \cite{Cvetic:2005zi} and the solution is pathological. On the other hand, naked CTCs can be avoided when \eqref{SUSY} holds and for this reason it is this condition that we will use in the remainder of this note.  

The supersymmetric values of the conserved charges are
\begin{align}
	E \ =& \ \frac{1}{G_N} \frac{1}{g} \frac{m \pi^2}{8} \frac{243 ~e^{-2 \delta } \left(e^{2 \delta }-1\right)^6 \left(-21 e^{4 \delta }+18 e^{6 \delta }+7\right) }{\left(3 e^{2 \delta } - 5\right)^4 \left(3 e^{2 \delta }-1\right)^3} \, , \label{EnSusy} \\
	J \ =& \ \frac{1}{G_N} \frac{m \pi^2}{8g} \frac{81 ~e^{-2 \delta } \left(e^{2 \delta }-1\right)^6 \left(18 e^{2 \delta }+9 e^{4 \delta }-23\right)}{\left(3 e^{2 \delta } - 5\right)^4 \left(3 e^{2 \delta }-1\right)^3} \, , \label{JSusy} \\
	Q \ =& \ \frac{1}{G_N} \frac{1}{2g} \frac{m \pi^2}{8} \frac{729~ e^{-2 \delta } \left(e^{2 \delta }-1\right)^7 \left(e^{2 \delta }+1\right) 
   }{\left(3 e^{2 \delta } - 5\right)^3 \left(3 e^{2 \delta }-1\right)^3} \, , \label{QSusy}
\end{align}
and satisfy
\begin{align}
	E - 3 J - 4 Q \ = \ 0 \, . \label{SusyReln}
\end{align}
The remaining quantities, such as the temperature, can also be evaluated in the supersymmetric limit but the resulting expressions are not compact so we do not present them here. However, the temperature is non-vanishing and consequently, as we will see in the next section, this means the supersymmetric black hole is not necessarily extremal.

\subsection{Extremality}

The black hole is extremal if the outer horizon coincides with another horizon. Since $Y(r)$ is a quartic function of $r^2$ we expect there to be four distinct horizons in general. We denote the location of these horizons by $(r_+,r_0,\tilde{r}_0,r_-)$ where $r^2_+ \geq r^2_0 \geq \tilde{r}^2_0 \geq r^2_-$. We may write
\begin{align}
	( g^2 \Xi^4 )^{-1} Y(r) \ =& \ ( g^2 \Xi^4 )^{-1} ( y_0 + y_1 r^2 + y_2 r^4 + y_3 r^6 + y_4 r^8 ) \\
	\equiv &\ ( r^2 - r^2_+ ) ( r^2 - r^2_0 ) ( r^2 - \tilde{r}^2_0 ) ( r^2 - r^2_- ) \, ,
\end{align}
so that the extremal limit, reached when $r^2_+=r^2_0$, corresponds to a double root of $Y(r)$.
Comparing coefficients determines
\begin{align}
	( g^2 \Xi^4 )^{-1} y_4  \ =& \ 1 \, , \\
	( g^2 \Xi^4 )^{-1} y_3 \ =& \ -r_+^2-\tilde{r}_0^2-r_-^2-r_0^2  \, , \\
	( g^2 \Xi^4 )^{-1} y_2  \ =& \ r_+^2 \left(\tilde{r}_0^2 + r_-^2 + r_0^2 \right) + r_0^2 \tilde{r}_0^2 + r_-^2 \tilde{r}_0^2 + r_-^2 r_0^2 \, , \\
	 ( g^2 \Xi^4 )^{-1} y_1  \ =& \ r_+^2 \left(-r_0^2 \tilde{r}_0^2 - r_-^2 \tilde{r}_0^2 - r_-^2 r_0^2 \right) - r_-^2 r_0^2 \tilde{r}_0^2 \, ,\\
	( g^2 \Xi^4 )^{-1} y_0 \ =& \ r_+^2 r_-^2 r_0^2 \tilde{r}_0^2 \;.
\end{align}
We now show that the double root $r_+^2=r_0^2$ also corresponds to a zero-temperature solution. Recall that the temperature is given by
\begin{align}
	T \ =& \ \left. \frac{\partial_r Y}{4\pi g \rho^3 \sqrt{\Xi f_1}} \right|_{r=r_+} \, ,
\end{align}
where
\begin{align}
	\partial_r Y |_{r=r_+} \ =& \ \frac{1}{r_+} ( 8 y_4 r^8_+ + 6 y_3 r_+^6 + 4 y_2 r_+^4 + 2 y_1 r_+^2 ) \, .   
\end{align}
Substituting for the $y$'s leads to
\begin{align}
	 T \ = \ \frac{g^2 \Xi^4}{4\pi g r_+ \rho^3 \sqrt{\Xi f_1(r_+)} } \big[ 2 r^2_+(r^2_+ - r^2_0)(r^2_+ - \tilde{r}^2_0)(r^2_+ - r^2_-) \big] \, ,
\end{align}
so that the temperature vanishes when $r_0$ coincides with $r_+$, i.e.\ in the extremal limit. Note that we have not used the supersymmetry condition and hence the extremal black hole is not necessarily supersymmetric.

\subsection{BPS Limit}

As in \cite{Cabo-Bizet:2018ehj} we will work with supersymmetric black-holes and study their behaviour in the extremal limit. When that happens, we use the nomenclature ``BPS'' (denoting both supersymmetric \emph{and} extremal solutions) and we label the corresponding quantities with a $*$.

The absence of naked CTCs is a physically sensible requirement and places further constraints on the parameters $(m,\delta,a)$ describing the black hole, rendering the supersymmetric solution extremal. One way of ruling out CTCs requires that \cite{Cvetic:2005zi}
\begin{align}
	0 \ =& \ H^{\frac{2}{5}} \left( - \frac{Y}{f_1} + \frac{f_1}{H^2 \Xi^2 \rho^4} \left( 2g + \frac{2f_2 \Xi_-}{f_1} \right)^2 \right) \, ,
\end{align}
at $r=r_+$. This can be achieved if, in addition to the SUSY condition \eqref{SUSY}, the following relation holds
\begin{align}
	m \ = \ m_* \ =& \ \frac{128 e^{2\delta} ( 3 e^{2\delta} - 1 )^3 }{729g^4 (e^{2\delta}+1)^2 (e^{2\delta}-1)^6 } \, . \label{mBPS}
\end{align}
When the parameter $m$ takes this value the function $Y$ has a double root as expected at $r_+^2=r_*^2 =r_0^2$ given by
\begin{align}
	r_*^2 \ =& \ \frac{16}{ 3g^2 (3e^{2\delta}-5) (e^{2\delta}+1)} \ = \ \frac{4a^2}{(1+ag)(1-3ag)} \, . \label{rBPS}
\end{align}
We may invert this expression to write the charge in terms of the BPS radius:
\begin{align}
	e^{2\delta} \ = \ \frac{1}{3} \left( 1 \pm  \frac{4\sqrt{g^2 r_*^2(1+g^2 r_*^2)}}{g^2 r_*^2}  \right) \, . \label{DeltarBPS}
\end{align}
Evaluating the thermodynamic quantities for these BPS values of $m$ and $r_+$ gives
\begin{align}
	T^* \ =& \ 0 \, , \label{TBps} \\
	S^* \ =& \ \frac{1}{G_N} \frac{\pi^3}{g^5} \frac{32 \sqrt{9 e^{2 \delta }-7}}{3 \sqrt{3} \left(3 e^{2 \delta }-5\right)^3 \left(e^{2 \delta }+1\right)^{3/2} } \, , \label{EntBps} \\	
	\Omega^* \ =& \ 1 \, , \label{OBps} \\
	\Phi^* \ =& \ 2 \, , \label{PBps} \\
	E^* \ =& \ \frac{1}{G_N} \frac{\pi^2}{g^5}\frac{16 \left(18 e^{6 \delta }-21 e^{4 \delta }+7\right)}{3 \left(3 e^{2 \delta } - 5\right)^4 \left(e^{2 \delta
   }+1\right)^2} \, , \label{EnBps} \\
   J^* \ =& \ \frac{1}{G_N} \frac{\pi^2}{g^5}\frac{16 \left(9 e^{4 \delta }+ 18 e^{2 \delta }-23\right)}{9 \left(3 e^{2 \delta } - 5\right)^4 \left(e^{2 \delta
   }+1\right)^2} \, , \label{JBps} \\
   Q^* \ =& \ \frac{1}{G_N} \frac{\pi^2}{g^5}\frac{8 \left(3 e^{6 \delta }-5 e^{4 \delta }-3 e^{2 \delta }+5\right)}{\left(3 e^{2 \delta }-5\right)^4 \left(e^{2 \delta }+1\right)^2 } \, . \label{QBps} 
\end{align}
The supersymmetry relation \eqref{SusyReln} with these expressions can be simply written as
\begin{align}
	E^* - 3 J^* \Omega^* - 2 Q^* \Phi^* \ = \ 0 \, . \label{SusyRelnBPS}
\end{align}

\subsection{Complexified Solution}\label{sec:26}

We can extract information about the value of the parameter $m=m_+$ at the outer horizon---but away from the extremal limit---by examining $Y(r_+)=0$. We see that $Y(r_+)=0$ is equivalent to a quadratic equation for $m_+$:
\begin{align}
	0 \ =& \ m_+^2 ( 4 g^2 s^4 )+ m_+ \Big[ 2 \left(g s^2 \left(2 g \Xi ^2 r_+^4-a^3 (a g+2)\right)-\Xi  r_+^2 \left(c^2 a^2
   g^2+\Xi \right)+a^2\right) \Big] \nonumber \\
	&\ + r_+^6 ( 1+ g^2 r_+^2 ) \Xi^4 \label{mQuad} \, .
\end{align}
Inserting the SUSY condition \eqref{SUSY}, we solve to find
\begin{align}
	m_+ \ =& \ -\frac{2 e^{2 \delta } \left(3 e^{2 \delta }-1\right) \left(\left(3 e^{2 \delta }-5\right) g^2 r_+^2 \left(9 e^{2 \delta }
   \left(\left(e^{2 \delta }-2\right) g^2 r_+^2-2\right)+5 g^2 r_+^2-2\right)+32\right)}{81 \left(e^{2 \delta }-1\right)^6 g^4} \nonumber \\
   &\ \pm \frac{2e^{2 \delta } \left(3 e^{2 \delta }-1\right) \left(16-3 \left(e^{2 \delta }+1\right)
   \left(3 e^{2 \delta }-5\right) g^2 r_+^2\right)}{81 \left(e^{2 \delta }-1\right)^6 g^2} \sqrt{\left(9 e^{2 \delta } \left(e^{2 \delta }-2\right)+5\right) g^2 r_+^2-4} \, .
\end{align}
As shown in App.~\ref{AppA}, the parameter $m_+$ is complex when $r_+>r_*$ and becomes real (and equal to its BPS value $m_*$) only at $r_+=r_*$.
Consequently, the thermodynamic quantities in which $m_+$ appears are generically complex away from the BPS limit. Substituting for this complex $m_+$ and $r_+$ gives cumbersome expressions for the thermodynamic quantities, which however can be shown to satisfy the remarkable condition:
\begin{align}
	2 \Phi - 3 \Omega - 1 \ = \ \pm  2 \pi i T \, . \label{Constraint}
\end{align}
Using this, one can show that the free energy \eqref{QSR} can be expressed very simply in terms of
\begin{align}\label{OSA}
	I \ =& \ - \frac{ \pi^2 }{128 g^5 G_N} \frac{\phi^4}{\omega^3} \, ,
\end{align}
where
\begin{align} 
  \label{eq:21}
2 \phi - 3\omega \ = \ \pm 2 \pi i  \;.
\end{align}
Here
\begin{align}\label{newchem}
	\omega \ =& \ \beta ( \Omega - \Omega^* ) \\
	\phi \ =& \ \beta ( \Phi - \Phi^* ) \,.
\end{align}

In terms of these redefined chemical potentials, the quantum statistical relation \eqref{QSR} becomes
\begin{align}
  \label{eq:20}
  I \ = \ - S - 3\omega J - 2 \phi Q\;, 
\end{align}
subject to the condition \eqref{eq:21}, where the energy has disappeared using the relation \eqref{SusyReln}. One could formally re-instate it by writing
\begin{align}
  \label{eq:22}
  I \ = \ \beta (E -3 \Omega^* J - 2 \Phi^* Q ) - S - 3\omega J - 2 \phi Q\;.
\end{align}
This form of the on-shell action will be useful shortly when establishing the background at the boundary of AdS$_7$.

The results \eqref{OSA} and \eqref{eq:21} obtained here for the free
energy reproduce those first written down in \cite{Hosseini:2018dob} following \cite{Hosseini:2017mds}. One sees from \eqref{eq:20} that a Legendre transform with respect to the chemical potentials $\omega$ and $\phi$ will yield the Bekenstein-Hawking entropy for the AdS$_7$ black hole. 

\subsection{SCFT Background from Bulk Regularity}

We will now shift our focus to recovering the form of the background at the conformal boundary located at $r=\infty$.

{\bf Metric}: We begin by looking at the form of the black-hole solution \eqref{BHmetric}-\eqref{BHscalar} in the limit $r\to\infty$. One  obtains
\begin{equation}\begin{split}
  \dd s^2_7 \ &= \ -g^2 r^2 \dd t^2 + r^2 (\sigma + g \dd t)^2 + \frac{1}{g^2 r^2}\dd r^2 + r^2 \dd s^2_{\mathbb{CP}^2}\\
  \ &= \ \frac{1}{g^2 r^2}\dd r^2 + r^2 \dd s^2_{\text{bdry}}\;,
\end{split}\end{equation}
with the boundary metric being
\begin{equation}\begin{split}
  \dd s^2_{\text{bdry}} \ =& \ -g^2 \dd t^2 + (\sigma + g \dd t)^2 + \dd s^2_{\mathbb{CP}^2}\\
  \to&\ -\dd t^2 + \sigma^2 + \dd s^2_{\mathbb{CP}^2} \\\
  \ =& \ -\dd t^2 + \dd s^2_{S^5}\;.
\end{split}\end{equation}
In the second line above  we performed a scaling of the time coordinate $t \to t/g$ and also absorbed $g\dd t$ into $\sigma$ by sending the fibre coordinate $\psi \to\psi - g \dd t$. The  boundary metric is therefore just  $\mathbb R \times S^5$ and does not depend on the chemical potentials defined in the bulk. This dependence will instead be recovered by looking at the behaviour of the metric at the horizon.

In order to do so, we first analytically continue to Euclidean gravity by letting $t = -i\tau$. One then introduces into \eqref{BHmetric} the shifted radial coordinate $R^2 = r - r_{+}$. At the horizon, the black-hole metric takes the following form
\begin{equation} \label{two}
  \dd s^2_7 \ = \ h_{RR}\left(\dd R^2 + R^2\left(\frac{2\pi}{\beta}\dd \tau\right)^2\right) + (r_+^2+h_{\sigma})\left(\sigma - \frac{2f_2}{f_1} i \dd \tau \right)^2 + r_{+}^2 \dd s^2_{\mathbb{CP}^2}\;,
\end{equation}
where $h_{RR}$ and $h_{\sigma}$ are functions of the parameters of the original metric and of some angular coordinates on $\mathbb{CP}^2$, the explicit form of which is not important for the ensuing analysis. By employing similar rescaling transformations as on the boundary metric, $\tau \to \tau/g$ and $\psi \to \psi- i\tau$ and using the definition of $\Omega$ from \eqref{Omega} we can re-write
\begin{equation}
  (r_+^2 + h_{\sigma})(\sigma + i \Omega \dd \tau)^2 + r_+^2 \dd s^2_{\mathbb{CP}^2} \ = \ 2i r_+^2 \Omega \sigma \dd \tau- r_+^2 \Omega^2 \dd \tau^2 +h_{\sigma}(\sigma + i\Omega \dd \tau)^2 + r_+^2 \dd s^2_{S^5}\;.
\end{equation}
At this stage we perform the coordinate transformation detailed in App.~\ref{AppB},\footnote{This coordinate transformation also brings the Killing vector \eqref{eq:16} to the form $\ell = \partial/\partial t + \Omega \sum_{i=1}^3 \partial/\partial \phi_i$.} which brings the metric on $S^5$ as well as the 1-form $\sigma$ to 
\begin{equation}\label{usingApp}\begin{split}
    \dd s_{S^5}^2 \ =& \ \dd \theta_1^2 + \sin^2\theta_1 \dd \theta_2^2 + \sin^2\theta_1\sin^2\theta_2 \dd \phi_1^2 + \sin^2\theta_1\cos^2\theta_2 \dd\phi_2^2 + \cos^2\theta_1 \dd\phi_3^2\;,\\
    \sigma \ =& \ \sin^2\theta_1\sin^2\theta_2 \dd\phi_1 + \sin^2\theta_1\cos^2\theta_2 \dd\phi_2 + \cos^2\theta_1 \dd\phi_3\;.
\end{split}\end{equation}
Using this fact, we can recast the metric close to the horizon as
\begin{align} \label{four}
  \dd s^2 \ = \ & h_{RR}\left(\dd R^2 + R^2\left(\frac{2\pi}{\beta}\dd \tau\right)^2\right) + h_{\theta_1\theta_1}\dd \theta_1^2 + h_{\theta_2\theta_2}\dd\theta_2^2 + h_{\phi_1\phi_1}(\dd\phi_1 + i\Omega \dd\tau)^2 \nonumber \\
  &+ h_{\phi_2\phi_2}(\dd\phi_2 + i\Omega \dd\tau)^2 + h_{\phi_3\phi_3}(\dd\phi_3 + i\Omega \dd\tau)^2 + h_{\phi_1\phi_2}(\dd\phi_1 + i\Omega \dd\tau)(\dd\phi_2 + i\Omega \dd\tau) \nonumber \\
  &+ h_{\phi_1\phi_3}(\dd\phi_1 + i\Omega \dd\tau)(\dd\phi_3 + i\Omega \dd\tau) + h_{\phi_2\phi_3}(\dd\phi_2 + i\Omega \dd\tau)(\dd\phi_3 + i\Omega \dd\tau) \;,
\end{align}
where we once again emphasise that the explicit form of the functions $h$ is not important for the remaining discussion.

The metric in Eq.~\eqref{four} describes a warped fibration of $S^5$ (parametrised by the coordinates $(\theta_1,\theta_2,\phi_1,\phi_2,\phi_3)$) over $\mathbb{R}^2$ (parameterised by $R$ and $\tau$). To ensure the lack of conical singularities at the point $R = 0$, one has to introduce the twisted identifications of certain coordinates.
\begin{equation}\label{newcoords}
  (\tau\;, \phi_1\;, \phi_2\;, \phi_3) \ \sim  \ (\tau+ \beta\;,\phi_1 - i\Omega \beta\;, \phi_2 - i\Omega \beta\;, \phi_3 - i\Omega \beta) \, ,
\end{equation}
as one completes a circle around the temporal direction. It is important to point out that the 1-form $i\Omega \dd\tau$ is not well defined at $R = 0$, but we can remove this dependence from the metric by introducing a ``rotating" coordinate frame:
\begin{equation}
\tau \ = \ \hat{\tau},\quad \phi_1 \ = \ \hat{\phi}_1 - i\Omega \hat{\tau},\quad \phi_2 \ = \  \hat{\phi}_2 - i\Omega \hat{\tau},\quad \phi_3 \ = \ \hat{\phi}_3 - i\Omega \hat{\tau}\, .
\end{equation}
We have thus ``untwisted'' the identifications to recover more canonical ones as we go around the Euclidean time circle
\begin{equation}\label{newercoords}
  (\hat{\tau}\;, \hat{\phi}_1\;,\hat{\phi}_2\;, \hat{\phi}_3) \ \sim \ (\hat{\tau} + \beta\;, \hat{\phi}_1\;, \hat{\phi}_2\;, \hat{\phi}_3)\;.
\end{equation}
We then take the
coordinates \eqref{newercoords} and substitute them into the boundary metric. This results in the following final form 
\begin{align}
  \dd s_{\text{bdry}}^2 \ =& \ \dd\hat{\tau}^2 + \dd\theta_1^2 + \sin^2\theta_1 \dd\theta_2^2 + \sin^2\theta_1\sin^2\theta_2 (\dd\hat{\phi}_1 - i\Omega \dd\hat{\tau})^2 \nonumber \\
  &\ + \sin^2\theta_1\cos^2\theta_2 (\dd\hat{\phi}_2 - i\Omega \dd\hat{\tau})^2 + \cos^2\theta_1 (\dd\hat{\phi}_3 - i\Omega \dd\hat{\tau})^2\;. \label{finalboundary}
\end{align}

{\bf 1-form gauge fields:} Next, we will address the remaining fields in the supergravity multiplet, switching momentarily back to Lorentzian signature. The only fields which are non-trivial at the conformal boundary are the 1-form gauge fields which become 
\begin{align}
	 \ A|_{\rm bdry} \ = \ \alpha \dd t \, ,
\end{align}
that is, they can only have a pure-gauge dependence. These terms cannot be fixed just by requiring regularity of the bulk metric at the horizon. However, they can be restricted by looking at the action of the Lie derivative with respect to the Killing vector \eqref{eq:16} on the Killing spinors \cite{Cabo-Bizet:2018ehj}.

The solutions to the Killing spinor equations for the backgrounds \cite{Chong:2004dy,Chow:2007ts}, or even the special case that we are considering with 2-equal charges and 3-equal rotations are not known. However, the Killing spinors for the background with 2 independent charges and vanishing rotations were given in \cite{Liu:1999ai}. For equal charges, this is a subcase of the configuration we are considering with $\Omega = 0$. Fortunately, it is also precisely what we need to fix the asymptotic form of the 1-form gauge fields at the boundary.

When the two charges are set to the same value, the Killing spinor given in \cite{Liu:1999ai} is schematically of the form
\begin{align}\label{KSML}
  \epsilon \ = \ e^{\frac{1}{4}g(1 + 2 \alpha)t\Gamma^{12}}(\dots)\mathcal{P} \epsilon_0\;,
\end{align}
where the ellipsis represents the angular and radial terms which commute with $\Gamma^{12}$, the $\mathcal{P}$ is a projection operator which also commutes with $\Gamma^{12}$ and $\epsilon_0$ is a constant spinor.\footnote{We note that our normalisations are slightly different when compared to \cite{Liu:1999ai}, $g_{\rm LM} = 4g$, $\Gamma^{12}_{\rm LM} = \frac{1}{2} \Gamma^{12}$. We have also added a pure-gauge term in the 1-form gauge fields.} The rank~2 $SO(5)$ Gamma matrix $\Gamma^{12}$ is such that the Killing spinors have the following eigenvalues, $\frac{i}{2}\Gamma^{12} \epsilon=\pm \epsilon$. 

With this information at hand, we proceed with the evaluation of the Lie derivative. For vanishing rotations the Killing vector \eqref{eq:16} simplifies to $\ell = \frac{\partial}{\partial t}$ and hence we simply need to evaluate $\mathcal{L}_{\frac{\partial }{\partial t}} \epsilon$. Explicit calculation shows that the Lie derivative simplifies dramatically 
\begin{align} \label{GKeq:2}
  \mathcal{L}_{\frac{\partial }{\partial t}} \epsilon \ = \ \frac{\partial \epsilon}{\partial t} \ = \ \mp \frac{i}{2}(1 + 2 \alpha)g\epsilon\;.
\end{align}
Keeping in mind that the electrostatic potential is defined as
\begin{align}
  \Phi \ =  \ \imath_\ell A|_{r_+} - \imath_\ell A|_{\rm bdry} = \imath_\ell A|_{r_+} - \alpha \, ,
\end{align}
with $\imath$ the interior product, one can define the gauge parameter as
\begin{align}
  \alpha \ \equiv \ \imath_\ell A|_{r_+} - \Phi \, .
\end{align}
By also using \eqref{Constraint}, Eq.~\eqref{GKeq:2} can finally be written as
\begin{align}
  \mathcal{L}_{\partial/\partial t} \epsilon \ = \ \mp \left(-\frac{\pi}{\beta} + i\imath_\ell A|_{r_+} \right)g\epsilon\;.
\end{align}
Analytically continuing to the Euclidean-signature solution, and using the coordinates \eqref{newercoords}, the circumference of the time circle is $\beta$. Transporting the Killing spinor around the time circle generated by $\ell$ can be done through the exponentiation of the action of the Lie derivative
\begin{align} \label{GKeq:4}
  e^{i \beta \mathcal{L}_{i\partial/\partial \hat{\tau}}} \epsilon \ = \ -e^{\pm\beta\imath_\ell A|_{r_+}g}\epsilon\;.
\end{align}
In order to keep the gauge field well defined, the component in the direction which shrinks as we go to the black-hole horizon has to vanish, which sets $  A|_{r_+} = 0$, while in order to satisfy \eqref{GKeq:4} the Killing spinor has to be \emph{anti-periodic} when transported all the way along the time circle. This discussion also fixes the pure gauge parameter once and for all to be
\begin{align}
  \label{eq:19}
 \alpha \ = \ -\Phi\; ,
\end{align}
and leads to the pure-imaginary gauge field at the boundary 
\begin{equation}
  \label{gaugefield}
  A|_{\rm bdry} \ = \ i \Phi \dd \hat{\tau}\;.
\end{equation}
Note that although we carried out this analysis for the charged AdS$_7$ black-hole solution with no rotations, the results should extend straightforwardly when one turns the rotations back on, as in \cite{Cabo-Bizet:2018ehj}.\footnote{For example, rotation is supported by the non-trivial 3-form potential in \eqref{BH3form} which does not have a component along the time direction and its presence only affects the $(\dots)$ part of the Killing spinor solution \eqref{KSML} with the relation \eqref{GKeq:2} being unchanged.} We will henceforth go back to considering the case with 3-equivalent rotations.

{\bf Energy:} Translations with respect to the new Euclidean time coordinate $\hat{\tau}$ will have a corresponding charge $\hat{E}$, given by 
\begin{equation}
  \label{eq:12}
  \hat{E} \ = \ E - 3 \Omega J \;.
\end{equation}
Utilising  this in the quantum-statistical relation \eqref{eq:22}, when formally also including the gauge field for which $  A|_{r_+} = 0$, leads to
\begin{align}
  \label{eq:25}
    I \ = \ \beta \left(\hat{E} +3(\Omega- \Omega^*) J + 2 (A|_{r_+}-\Phi^* )Q \right) - S - 3 \omega J - 2 \phi Q\;,
\end{align}
where the term multiplying $\beta$ is zero via \eqref{SusyReln}. However, at the boundary  this combination can be interpreted as a supersymmetric Hamiltonian, which can be further simplified to
\begin{equation}
  \label{eq:13}\begin{split}
  \{\QQ,\QQ^\dagger\} \ =& \ \hat{E} + 3(\Omega - \Omega^*)J + 2 ( A|_{\rm bdry} -
  \Phi^*) Q\\
  \ =& \ \hat{E} + \frac{1}{\beta}(3\omega J+ 2 \varphi Q)\;, \;
\end{split}\end{equation}
using \eqref{gaugefield} and the redefined chemical potentials \eqref{newchem}.

{\bf Summary}: To summarise, we have derived the following information about the supersymmetric, black-hole solution at the boundary of AdS$_7$: 
\begin{itemize}
\item[$\diamond$] It will involve the following metric
\begin{equation}\begin{split}
  \dd s_{\text{bdry}}^2 \ =& \ \dd\hat{\tau}^2 + \dd\theta_1^2 + \sin^2\theta_1 \dd\theta_2^2 + \sin^2\theta_1\sin^2\theta_2 (\dd\hat{\phi}_1 - i\Omega \dd\hat{\tau})^2 \\
  &\ + \sin^2\theta_1\cos^2\theta_2 (\dd\hat{\phi}_2 - i\Omega \dd\hat{\tau})^2 + \cos^2\theta_1 (\dd\hat{\phi}_3 - i\Omega \dd\hat{\tau})^2\;,
\end{split}\end{equation}
which is a nontrivial fibration of $S^1_\beta$ over $S^5$.

\item[$\diamond$] There will be a background gauge field
  \begin{align}
   A|_{\rm bdry} \ = \ i \Phi \dd \hat{\tau}\;.    
  \end{align}

\item[$\diamond$] The Killing spinor will satisfy anti-periodic boundary conditions on  $S^1_\beta$. 

  \item[$\diamond$] At the boundary, the charge associated with translations $\partial/\partial \hat{\tau}$ is given by 
\begin{equation}\label{Ehat}
 \begin{split}
\beta \hat{E} \ = \ \beta  \{\QQ,\QQ^\dagger\} - 3\omega J - 2 \varphi Q\;.
\end{split}\end{equation}

\item[$\diamond$] The chemical potentials obey the constraint 
  \begin{align}\label{listedconstraint}
2 \phi - 3\omega \ = \ \pm 2 \pi i  \;.
\end{align}

\end{itemize}

\section{(2,0) Partition Function on the Boundary}\label{boundary}

We now shift our attention to the dual field theory. An extremisation principle that reproduces the bulk AdS$_7$ black-hole entropy was first proposed in \cite{Hosseini:2018dob}, based on an insightful modification of the Casimir-energy result of \cite{Bobev:2015kza}. Here, we use the bulk calculation of the preceding section to provide a microscopic derivation of \cite{Hosseini:2018dob}, in line with expectations from the approach of \cite{Cabo-Bizet:2018ehj}.

\subsection{Expectations from AdS/CFT}\label{indexcft}

The AdS/CFT correspondence dictates that at large $N$ the on-shell action for the supersymmetric AdS$_7$ black-hole solution \eqref{OSA} should match the generator of connected correlators in the boundary theory, that is $ I=-\log \ZZ$ \cite{Witten:1998qj}. Because of the anti-periodic boundary conditions on the Killing spinors, and assuming usual periodic boundary conditions for the bosons, all fermions will be anti-periodic. Let us encode this information by allowing for general $e^{\pi i n_0}$ periodicity the fermions in the boundary theory with the understanding that we need to set $n_0 = \pm 1$ when comparing with gravity. The partition function is then to be understood as
\begin{align}
  \label{eq:23}
  \ZZ \ \sim \ \Tr_{\HH} e^{- \beta \hat{E}}
\end{align}
since $\hat{E}$ corresponds to translations in the time variable $\hat{\tau}$ and where the trace is over the physical Hilbert space of the flat-space $A_{N-1}$ (2,0) theory in radial quantisation. Through \eqref{Ehat} it is clear that this is a special case of the index-like quantity
\begin{equation}
  \label{eq:2}
  \II(\omega_1, \omega_2, \omega_3, \phi_1, \phi_2; n_0 ) \ = \ \Tr_{\mathcal H } (-1)^{F(1+n_0)} e^{-\beta \{\QQ, \QQ^\dagger\}+ \beta \sum_{i=1}^3 \omega_i J_i + \beta\sum_{j=1}^2\phi_j Q_j }\;,
\end{equation}
where the chemical potentials are subject to the condition
\begin{align}
  \label{eq:30}
\sum_{i=1}^3 \omega_i  - \sum_{j=1}^2\phi_j
= \frac{2 \pi i n_0}{\beta}\;,\qquad n_0\in \mathbb Z\;.  
\end{align}
 In the above, $\QQ$ denotes any one of the Poincar\'e
supercharges preserved by the 6D theory, while the  $\{\Delta,
\omega_i\}$, $\{\phi_j\}$ are Cartan generators (dilatations,
orthogonal rotations and R-charges respectively)  for the maximal
bosonic subalgebra of the 6D superconformal algebra
$\mathfrak{so}(2,6)\oplus\mathfrak{sp}(2)\subset\mathfrak{osp}(8^*|4)$. For concreteness, since we will be using the conventions of \cite{Bobev:2015kza}, our chosen supercharge will have charges $Q_1 = Q_2 = \frac{1}{2}$ and $J_1 = J_2 = J_3 = -\frac{1}{2}$. The
spin-statistics theorem allows us to express the fermion-number
operator as $F = 2J_1$. Note that we have rescaled our chemical potentials as $\{\omega_i, \phi_j\}\to \beta \{\omega_i, \phi_j\}$.

On the one hand, for $n_0 = 0$, the above expression reproduces precisely the definition of the most general 6D superconformal index of \cite{Bhattacharya:2008zy}, as presented in \cite{Bobev:2015kza}. On the other hand, the case which is of interest to us---as dictated from the gravity calculation---is the one with $\omega_i = \omega$, $J_i = J$, $\phi_j = \phi$, $Q_j = Q$ and $n_0 = \pm 1$. Keeping that in mind, it will be possible to keep all parameters and charges generic for the duration of the following discussion and fix them only at the very end.

Before proceeding further, let us massage Eq.~\eqref{eq:2} by defining a new parameter $m$ through the relations\footnote{This paramater $m$ should not be confused with the parameter of the same name appearing in Sec.~\ref{bulk}. We hope that the repeated  use of this symbol will not cause confusion.}
\begin{equation}
  \label{eq:4}
\phi_1 \ \equiv \ \frac{1}{2}\sum_{i=1}^3\omega_i -m - \frac{\pi i n_0}{\beta}\;,\qquad  \phi_2 \ \equiv \ \frac{1}{2}\sum_{i=1}^3\omega_i +m - \frac{\pi i n_0}{\beta}\;.
\end{equation}
The quantity $\II$ can be rewritten as
\begin{equation}
  \label{eq:3}
  \II(\omega_1, \omega_2, \omega_3, m; n_0) \ = \ \Tr_{\mathcal H } (-1)^{F(1+n_0)} e^{-\beta \{\QQ, \QQ^\dagger\}+ \beta\sum_{i=1}^3 \omega_i (J_i + \frac{Q_1 + Q_2}{2})+\beta m \frac{Q_2 - Q_1}{2} - 2 \pi i n_0 \frac{Q_1 + Q_2}{2} }\;.
\end{equation}
Making use of $2J_1 = F$ leads to
\begin{equation}
  \label{eq:5}
  \II(\omega_1, \omega_2, \omega_3, m; n_0) \ = \ \Tr_{\mathcal H } (-1)^{F} e^{-\beta \{\QQ, \QQ^\dagger\}+ \beta(\omega_1 - \frac{2 \pi i n_0}{\beta}) (J_1 + \frac{Q_1 + Q_2}{2})+  \beta\sum_{i=2}^3 \omega_i (J_i + \frac{Q_1 + Q_2}{2})+\beta m \frac{Q_2 - Q_1}{2} }\;.
\end{equation}
At this stage one can redefine\footnote{By a different assignment of the fermion-number operator, the shifts can also be accommodated in the other chemical potentials. }
\begin{equation}
  \label{eq:6}
  \tilde{\omega}_1 \ \equiv \ \omega_1 - \frac{2 \pi i n_0}{\beta}\;,\qquad \tilde{\omega}_2 \ = \ \omega_2\;, \qquad \tilde{\omega}_3 \ = \ \omega_3\;.
\end{equation}
and write
\begin{equation}
  \label{eq:8}
  \II(\omega_1, \omega_2, \omega_3, m; n_0) \ = \ \II(\tilde{\omega}_1, \tilde{\omega}_2, \tilde{\omega}_3, m; 0)\;.
\end{equation}
Therefore, the quantity $\II$ can be thought of as a bona fide superconformal index when associated with the tilded chemical potentials, i.e. it is independent of $\beta$. It also has an $N$-scaling of $O(1)$ as $N\to\infty$; see \cite{Cabo-Bizet:2018ehj} and also the nice discussion in \cite{Larsen:2019oll}.

Note that when expressed as a supersymmetric path integral, the partition function is expected to factorise into an expression of the form:\footnote{This is indeed the case for cases where the exact supersymmetric path integral can be evaluated directly.}
\begin{equation}
  \label{eq:1}
  \ZZ_{S^5 \times S^1_{\beta}} \ = \  e^{- \beta\FF} \II \;, 
\end{equation}
where $\mathcal F$ is referred to as the ``generalised supersymmetric Casimir energy''.\footnote{When $n_0=0$ this reduces to the supersymmetric Casimir energy of \cite{Kim:2012ava}.} Connecting to the gravitational result for the on-shell action requires that $\FF$ has a large-$N$ scaling of $O(N^3)$. The evaluation of the RHS for the above equation at large $N$ will be our next goal.

Although analogous partition functions for superconformal theories in three and four dimensions can be directly computed using supersymmetric localisation, this is not the case for the interacting (2,0) theory, which lacks a Lagrangian description. However,  for the case where $n_0 = 0$ heroic technical works \cite{Kallen:2012cs,Kallen:2012va,Kallen:2012zn,Kim:2012ava,Kim:2012qf,Kim:2013nva}  produced a candidate 6D result from the $S^5$ partition function of $SU(N)$ 5D MSYM.

Here we will base our calculation on the results of this partition-function calculation at large $N$ following \cite{Bobev:2015kza}, while deferring comments about the choice of 5D theory to Sec.~\ref{implications}. The key observation is that the indices in \eqref{eq:8} are expected to be reproduced by the above partition-function calculations in 5D when using the tilded chemical potentials. The answer turns out to be indeed of the form \eqref{eq:1} where the quantity $\FF$ gives back the exact value for the on-shell action predicted by AdS/CFT {\it including the overall coefficient}, when we take the Cardy-like limit $\omega_i \to 0$.

\subsection{The $S^5$ Partition Function at Large $N$ and a Cardy-like Limit}\label{5Dderiv}

The parameters of the 5D partition function on the (squashed) $S^5$ can be straightforwardly inherited from the 6D background corresponding to the RHS of \eqref{eq:8}, via dimensionally reducing on $S^1_\beta$ \cite{Imamura:2012bm}. The $\omega_i$ are associated with the squashing parameters of the five sphere while $m$ plays the role of a  mass parameter. The relationship between the (2,0) theory and 5D MSYM also identifies the radius of the thermal circle with the 5D gauge coupling, $2 \pi \beta = g_{\rm YM}^2$, with these quantities defined in units of the $S^5$ radius. The 6D physics are then expected to be recovered in the $\beta\to\infty$ limit, where the M-theory circle can no-longer be neglected.

As argued in \cite{Bobev:2015kza},\footnote{This argument relies on the results of \cite{Jafferis:2012iv}.} to leading order in the large $N$ limit the partition function on $S^5$ should not receive instanton contributions. This argument significantly simplifies the calculation, as the leading $N$ result in supersymmetric localisation only receives classical contributions from the localising saddle point and one-loop determinant contributions from the $\NN=1$ vector multiplet and single adjoint $\NN=1$ hypermultiplet.

These one-loop determinants in 5D can be obtained following the procedure of \cite{Imamura:2012bm}; see also \cite{Lockhart:2012vp,Kim:2012qf}. In that reference, the starting point was precisely \eqref{eq:2} in 6D for $n_0 = 0$. The one-loop determinants for 6D $(1,0)$ vector and hypermultiplets were reduced on the thermal circle to produce the 5D $\NN = 1$ contributions in terms of triple-sine functions, $S_3(x)$. When used in the 5D MSYM calculation of \cite{Bobev:2015kza}, for one vector and one adjoint hypermultiplet, the result is \begin{align}
  \label{eq:28}
    \ZZ_{\rm 1-loop}^{n_0 = 0} \ = \ \left(\frac{\text{lim}_{x \rightarrow 0} S_3(x)/x}{S_3(\tilde{m})}\right)^{N-1} \prod_{i > j}^N \frac{S_3(\pm i\lambda_{ij}|\vec{\omega})}{S_3(\tilde{m}\pm i\lambda_{ij}|\vec{\omega})}\;,
\end{align}
where we are using standard shorthand notation with $S_3(\pm x | \vec \omega) = S_3( x | \vec \omega) S_3(- x | \vec \omega)$, the equivariant mass is defined as $\tilde{m} = m + \frac{1}{2}(\omega_1+\omega_2+\omega_3)$ and $\lambda_{ij} = \lambda_i -\lambda_j$ are Coulomb-branch parameters. In this expression, the numerators encode fermion contributions from the vector multiplets while the denominators scalar contributions from the  hypermultiplets, after supersymmetric cancellations.

It is possible to revisit the derivation of \eqref{eq:28} from \cite{Imamura:2012bm} and repeat it for the case of general $n_0$. We show in App.~\ref{AppC} that this results in the following modification of the arguments in the triple-sine functions \eqref{eq:28}
\begin{align}
  \label{eq:29}
    \ZZ_{\rm 1-loop}^{n_0 } \ = \ \left(\frac{\text{lim}_{x \rightarrow 0} S_3(x)/x}{S_3(\frac{\omega_1 + \omega_2 + \omega_3}{2} - \frac{i \pi n_0}{\beta})}\right)^{N-1} \prod_{i > j}^N \frac{S_3(\pm i\lambda_{ij}|\vec{\omega})}{S_3(\frac{\omega_1 + \omega_2 + \omega_3}{2}- \frac{i \pi n_0}{\beta} \pm i\lambda_{ij}|\vec{\omega})}\;.
\end{align}

Combining this  with the classical contribution from the localising saddle-point, one obtains in the large-$N$ limit \cite{Bobev:2015kza}
\begin{equation}\label{5DPF}
  \ZZ^{n_0}_{S^5}(m,\vec{\omega}, \vec{\phi},\beta) \ = \ \frac{1}{(\omega_1 \omega_2)^{\frac{N-1}{2}}}\int \frac{\text{d}^{N-1}\lambda}{N!}e^{-\frac{2\pi^2 }{\beta\omega_1 \omega_2 \omega_3}\sum_i \lambda_i^2}   \ZZ_{\rm 1-loop}^{n_0 } (\lambda,\vec{\omega},\vec{\phi},\beta)\;,
\end{equation}
where the integration is over the Coulomb branch parameters $\lambda_i$; the scalar vev of the $\NN=1$ vector multiplet onto which the path integral localises. It takes values in the Cartan subalgebra of $SU(N)$.

We will now specialise to the case of interest with $\omega_1 = \omega_2 = \omega_3 = \omega$ and $\phi_1 = \phi_2= \phi$. The constraint hence reduces to $3 \omega - 2 \phi = \frac{2 \pi i n_0}{\beta}$. Using this we can write
\begin{align}
  \label{eq:31}
    \ZZ_{\rm 1-loop}^{n_0 } =   \left(\frac{\text{lim}_{x \rightarrow
  0} S_3(x)/x}{S_3(\phi)}\right)^{N-1} \prod_{i > j}^N \frac{S_3(\pm
  i\lambda_{ij}|\vec{\omega})}{S_3(\phi\pm i \lambda_{ij}|\vec{\omega})}
\end{align}
Performing the matrix integral of \eqref{5DPF} exactly is challenging. In the $n_0 = 0$ case the authors of \cite{Bobev:2015kza} employed the $\beta\to \infty$ limit to simplify the triple-sine functions. For general $n_0$ we will use an additional, Cardy-like limit in the spirit of \cite{Choi:2018hmj} by also considering $\omega\to 0$. We show in App.~\ref{AppC} that the 1-loop-determinant contribution is then approximated by
 \begin{equation}
   \label{eq:9}
   \begin{split}
     \ZZ_{\rm 1-loop}^{n_0 }\ \simeq -\frac{\pi}{\omega^3}\left[\phi^2\sum_{i>j}\lambda_{ij}+ O(\beta, \omega)\right]\;,
   \end{split} \end{equation}
when the parameters $\lambda$ are restricted to the Weyl chamber where $\lambda_i >\lambda_j$ for $i>j$ and  assumed to be very large while the other parameters remain of order one \cite{Bobev:2015kza}. The partition function then simplifies to 
\begin{align}
  \label{eq:32}
    \ZZ^{n_0}_{S^5}(m,\vec{\omega},\beta) \ \propto \ \int \frac{\text{d}^{N-1}\lambda}{N!}e^{-\frac{2\pi }{ \omega^3} f(\lambda,\phi,\beta)}\;,
\end{align}
where
\begin{align}
  \label{eq:33}
  f = \frac{\pi}{\beta}\sum_i^N \lambda_i^2 + \frac{\phi^2}{2}\sum_{i>j}\lambda_{ij} + O(\beta,\omega)\;.
\end{align}
The integration in \eqref{eq:32} can now be carried out using the saddle-point approximation. Note that in our Cardy-like limit the constraint has reduced to $\phi = \frac{\pi i n_0}{\beta}$, which implies that $\phi^2 <0$. This is crucial to ensure that there exists a saddle point---the $\lambda_i$ are already ordered. The solution is \cite{Bobev:2015kza}
\begin{equation}
  \lambda_i \ = \ -\frac{\beta \phi^2}{4 \pi }(2i - N - 1)\;,
\end{equation}
which when substituted back into the integral leads to the following leading-$N$ expression for the 5D free energy
\begin{equation}\label{logZ}
  -\log \ZZ_{S^5} \ \propto\ -\frac{\beta \phi^4 N^3}{24 \omega^3} + O(\beta^0,\omega^{-2},N)\;.
\end{equation}
The claim is that this 5D calculation captures precisely the 6D partition function of the (2,0) theory on $S^5 \times S^1$, Eq.~\eqref{eq:1}, and hence that
\begin{align}
  \label{eq:24}
  -\log \ZZ_{S^5} \ = \ \beta \FF - \log \II\;.
\end{align}
In certain limits of the chemical potentials where finite-$N$ calculations can be explicitly performed and the $n_0 = 0$  superconformal-index can be evaluated independently, one sees that the corrections in \eqref{logZ} capture $O(N)$ terms in the quantity $\FF$ (e.g.\ coming from instantons) as well as the correct superconformal index contributions $\II$ which are $O(\beta^0, N^0)$ \cite{Bhattacharya:2008zy,Kim:2012ava,Kim:2012qf,Bullimore:2014upa,Beem:2014kka}.

In the large $N$ limit  therefore the expectation is that 
\begin{equation}
  \label{eq:10}
  -\log{\ZZ_{S^5 \times S^1}} \ \xrightarrow[\lim_{N\to \infty}] \ \beta\FF \;
\end{equation}
with
\begin{equation}
  \label{eq:11}
 \beta \FF \ = \ -\frac{\beta N^3\phi^4}{24 \omega^3} + O(N)\;.
\end{equation}
After setting $n_0 = \pm1$ as well as converting to gravitational parameters using  $N^3 = \frac{3 \pi^2 }{16 g^5 G_N}$ we arrive at
\begin{equation}
  \label{eq:7}
  \beta\FF \ = \ -\frac{\beta\pi^2 }{ 128g^5 G_N}\frac{\phi^4}{ \omega^3} + O(\beta^0, \omega^{-2},N)\;. 
\end{equation}
Upon sending $\{\omega,\phi\}\to\beta^{-1}\{\omega,\phi\}$,  this expression, which is valid in the Cardy-like limit $\omega\to 0$, exactly reproduces the gravitational on-shell action \eqref{OSA} and therefore yields the value of the black-hole entropy in AdS$_7$ with the correct normalisation.

\section{Implications for the Relation between 5D MSYM and the 6D (2,0) Theory}\label{implications}

Let us summarise our findings. We have shown that, following the analysis of \cite{Cabo-Bizet:2018ehj}, the AdS/CFT dictionary predicts a particular value for the generalised supersymmetric Casimir energy in the dual six-dimensional field theory. This value can then be reproduced from a five-dimensional partition-function calculation on (a squashed version of) $S^5$, once an appropriate identification of parameters is made. In this section we will initiate an investigation about the implications of the AdS/CFT matching with regards to the five-dimensional field theory used for supersymmetric localisation.

\subsection{Possible Ambiguities in the $S^5$ Partition Function }\label{assumptions}

The evaluation of the full 5D supersymmetric partition function using supersymmetric localisation is technically involved. From a five-dimensional viewpoint, there are several choices and potential ambiguities that could change the overall result. Here, we will list the ingredients responsible for reproducing the (generalised) supersymmetric Casimir energy Sec.~\ref{boundary}. \begin{enumerate}[(i)]
  
\item{Choice of theory:} Our prescription for extracting the Casimir energy involved using the two-derivative 5D MSYM in the $\beta\to \infty $ regime.

\item{Choice of background:} The background geometry for the 5D calculation is determined by the 6D superconformal index \eqref{eq:8} by dimensional reduction. In turn, the latter is fixed by the boundary behaviour of the AdS$_7$ black-hole solution, as discussed in Sec.~\ref{bulk}.
  
\item{Constant shifts:} It is possible to add constant shifts to the 5D action involving negative powers of $\beta$ and positive powers of curvature invariants without breaking any of  the symmetries. Similar terms appear in the partition function for the twisted 4D $\NN=4$ SYM theory on $S^4$ and fixing them is essential to ensure S-duality invariance \cite{Kim:2012ava,Kim:2012qf}. However, such contributions will be subleading in the $\beta\to \infty$ limit that we use to recover the Casimir energy.
  
\item{Choice of regularisation:} The 1-loop determinants require careful regularisation and this choice does affect the value of the exponential in \eqref{eq:1}. In \cite{Kim:2012ava,Kim:2012qf,Kim:2013nva} a regularisation that respects supersymmetry was employed and it was argued that this reproduces the directly-evaluated supersymmetric Casimir energy---that is $\langle \hat{E}\rangle$. For special limits of the parameters (e.g.\ when the 5D theory is placed on the round $S^5$ and for a specific value of the hypermultiplet mass)  maximal supersymmetry is preserved and this agreement can be explicitly checked using localisation. For a thorough discussion of this point see \cite{Kim:2016usy,Pestun:2016zxk}.
  
\item{Instantons subleading at large $N$:} In Sec.~\ref{boundary} an assumption was made about instanton contributions to the 5D free energy being subleading at large $N$. General arguments in support of this have been given in \cite{Jafferis:2012iv}.\footnote{Explicit examples of matching perturbative partition functions for $\mathcal{N}=1$ $USp(2N)$ gauge theories with corresponding 6D dual on-shell actions have been given in \cite{Alday:2014rxa,Alday:2014bta,Alday:2015jsa}.} In our context, at the maximally-supersymmetric point in chemical-potential space the instanton series can be resummed. It can then be explicitly checked that the instanton-sector contributions to the (generalised) supersymmetric Casimir energy are indeed subleading. 

 \end{enumerate}
Note that, while (ii)-(v) are well motivated assumptions by either physical or technical considerations, the important choice of theory (i) seems rather ad hoc: As 5D MSYM is perturbatively non-renormalisable \cite{Bern:2012di}, one could expect a host of higher-derivative interactions to be important in the $\beta\to\infty$ limit. Let us expand upon this point.\footnote{We remind the reader that compactifying the (2,0) on a circle leads relates $\beta \sim g_{YM}^2$.} 

An original motivation for the choice (i) was the conjecture of \cite{Douglas:2010iu,Lambert:2010iw,Papageorgakis:2014dma}. The claim in those works is that perturbative 5D MSYM plus instanton-charged solitons---including ``small'' solitons---contains all information about the physics of the (2,0) theory on a circle without the need to add any higher-derivative corrections. The status of this conjecture is not settled: Even though ``small'' instanton-solitons are not included in the perturbative definition of 5D MSYM, and should probably better be thought of as additional degrees of freedom,\footnote{Finite-sized instanton-solitons are expected to be exponentially suppressed as usual in perturbative amplitudes \cite{Papageorgakis:2014dma}.} there is no known mechanism for how these could cancel the expected perturbative UV divergences, such as the 6-loop divergence associated with the $ D^2\Tr F^4$ counterterm found in \cite{Bern:2012di}.\footnote{Note, however, that in the context of supergravity perturbative divergences can indeed be cancelled by running small black holes in loops using the framework of exceptional field theory \cite{Bossard:2015foa,Bossard:2017kfv}.}

A different, less radical justification for the ``unreasonable effectiveness'' of 5D MYSM has been the following: The result for the partition function is blind to $\QQ$-exact deformations, for any supercharge $\QQ$ that annihilates the action. It could be that higher-derivative interactions are all of that type and the calculation will therefore not depend on them \cite{Jafferis:2012iv}. However, a careful enumeration of the possible higher-derivative interactions---even at dimension 8---shows that this is not the case \cite{Bossard:2010pk,Chang:2014kma}.

\subsection{Higher-derivative Corrections to Localisation on $S^5$}

Supersymmetric higher-derivative corrections to 5D MSYM can be classified into D-terms and F-terms. The former are $\QQ$-exact and, as such, will not affect the result for the partition function; as a result  we will only focus on the latter. Of the F-terms, there can exist terms which are just $\QQ$-closed. The lowest-dimension such corrections are the supersymmetric completions of the $\frac{1}{2}$-BPS F-terms $\Tr F^4$, $(\Tr F^2)^2$ and the $\frac{1}{4}$-BPS F-term $D^2(\Tr F^2)^2$ with specific contractions of Lorentz indices \cite{Bossard:2010pk,Chang:2014kma}; for a concise summary see \cite{Mezei:2018url}. Note that the dimension-10 counterterm associated with the 6-loop divergence of 5D MSYM, the non-BPS operator $ D^2\Tr F^4$, is $\QQ$-exact (D-term) so our discussion has no bearing on this issue. 
Schematically, the form of the 5D effective action is given by
\begin{align}
  \label{eq:26}
  \LL_{\rm eff} \ = \ \LL_{\rm MSYM} + \LL_{ \QQ-{\rm closed}} +  \LL_{ \QQ-{\rm exact}}\;.
\end{align}
In the above
\begin{align}
  \label{eq:27}
\LL_{ \QQ-{\rm closed}} \ = \ \gamma_1 g_{YM}^8 \LL_{\Tr F^4} + \gamma_2 g_{YM}^8 \LL_{(\Tr F^2)^2} +\gamma_3 g_{YM}^{12} \LL_{D^2 (\Tr F^2)^2} + O(g_{YM}^{14})\;,
\end{align}
where $\LL_{\Tr F^4}$ indicates the maximally-supersymmetric completion of $\Tr F^4$ in 5D and so on, while the $\gamma_i$ denote undetermined coefficients.

It is straightforward to determine that for each higher-derivative correction of the form $\LL_{D^{2n} \Tr^k F^{2m}}$ the terms of interest in the localisation computation will be $\Tr^k \Phi^{2m}$.\footnote{In flat space the supersymmetric completion will include a factor $D \Phi$ in the adjoint scalar, for each power of  $F$. When put on the round $S^5$ each of these factors will be shifted to $D\Phi\to D\Phi + \frac{1}{r} \Phi$, where $r$ is the $S^5$ radius which we will now set to one. Modifying the classical action will not affect the localising saddle-point, which in the zero-instanton sector sets all fields to zero, including derivatives $D\Phi$,  except for the vev of $\Phi$\cite{Kim:2012ava}. One does have contributions from a component of an auxiliary field in the adjoint hypermultiplet but this will only change the overall coefficient in our discussion.} As a result, one can trace how the addition of higher-derivative terms in the action modifies the matrix model obtained through supersymmetric localisation. What changes, in particular, is the term corresponding to the classical action, i.e.\ the exponential term in \eqref{5DPF}.
One has
\begin{equation}
  e^{-\frac{2\pi^2}{\beta \omega_1 \omega_2 \omega_3}\sum_i \lambda_i^2} \longmapsto e^{-\frac{2\pi^2}{\beta \omega_1 \omega_2 \omega_3}\sum_i \lambda_i^2 + P(\beta,\{\lambda_i\})} \;,
\end{equation}
where $P$ is a polynomial in $\beta$ and the Coulomb-branch parameters $\lambda_i$, given by $ P(\beta,\{\lambda_i\}) \propto \beta^{2n+4m-5} \lambda^{2m}$, where the precise combination of $\lambda$s is given by the trace structure of the operator and up to an overall coefficient.

We will not attempt to solve this but there are some comments that one can immediately make. First, the $\beta$ scaling of the added terms is such that the simplifying assumptions in the derivation of the saddle-point equation used in Sec.~\ref{boundary} are no-longer valid. Second, it is clear that each successive addition will be increasingly-important in the large $\beta$ limit and therefore will significantly change the nature of the result. As a result, one would have to include {\it all} $\QQ$-closed corrections in \eqref{eq:27} before attempting to evaluate the matrix model.

\subsection{Summary of Checks for the 5D Partition Function}

To recapitulate, we emphasise that the 5D MSYM partition function on $S^5$ with the above assumptions can successfully  reproduce many features of the expected 6D partition function for the (2,0) theory in the form $\ZZ_{S^5 \times S^1} = e^{-\beta\FF} \II$ :
\begin{enumerate}[(a)]

\item It reproduces the 6D superconformal index $\II$ exactly in the abelian case where the latter can be calculated in terms of plethystic exponentials of free fields \cite{Bhattacharya:2008zy,Kim:2012ava,Kim:2012qf}.

\item An expansion of $\II$ in fugacities shows that in the nonabelian case the coefficients are integers and hence the result does indeed behave like an index \cite{Bhattacharya:2008zy,Kim:2012ava,Kim:2012qf}.

\item In a specific (Schur) limit of parameters $\II$ reproduces results for characters of protected chiral algebras, capturing the physics of certain BPS subsectors of the (2,0) theory \cite{Bullimore:2014upa,Beem:2014kka}.

  \item As shown in Sec.~\eqref{boundary}, $\FF$ reproduces the correct value for the supersymmetric Casimir energy, including the numerical coefficient.

\end{enumerate}

On the one hand that the results (a)-(c) pertain to $\II$ and are unaffected by the addition of higher-derivative interactions to 5D MSYM. On the other hand, the value of the generalised supersymmetric Casimir energy  $\FF$ can indeed change. Hence the precision AdS/CFT matching that we have established should significantly restrict the form of possible higher-derivative interactions.

The simplest way to implement this restriction would be to set the coefficients for all F-term higher-derivative corrections to zero. However, although we find it less likely, we cannot prove that there is no solution to the matrix model generated by including {\it all}  higher-dimension $\QQ$-closed operators, which also reproduces precisely the same value for the generalised supersymmetric Casimir energy. In any event, if that were the case, then the coefficients of the derivative expansion \eqref{eq:27} would also be uniquely fixed by the AdS/CFT duality.

\ack{ \bigskip We would like to thank E.~Andriolo, N.~Bobev, M.~Buican and
  especially S.~Murthy for many useful discussions and comments. We would also like to thank
  J.~Hayling for collaboration in the initial stages of this project and R.~Panerai for many enjoyable discussions, collaboration on related topics and comments on the manuscript. C.P.\ is
  supported by the Royal Society through a University Research
  Fellowship, grant number UF120032. P.R.\ is funded through the STFC grant ST/L000326/1.}

\newpage 
\begin{appendix}

\section{Reality Properties of $m$}\label{AppA}

Here we would like to show that the mass parameter appearing in the
AdS$_7$ black-hole metric \eqref{BHmetric} becomes complex away from the BPS limit (i.e.\ when $r_+ > r_*$) and becomes real when the black hole is supersymmetric and extremal (when $r_+ = r_*$).

We can check whether if any solutions of $m$ are complex by looking at
the discriminant
\begin{align}\label{discriminant}
  \Delta \ = \ C_1^2 - 4C_0 C_2\;,
\end{align}
with $C_i(r_+,r_*,a)$ being the coefficient of the $i$th power of $m_+$ in \eqref{mQuad}. The parameter $a$ can be expressed in terms of $r_*$ as
\begin{align} \label{one}
  a \ = \ \frac{-r_*^2 \pm 2\sqrt{r_*^2(1+r_*^2)}}{4+3r_*^2}\;,
\end{align}
where we have chosen to set the AdS radius to one, $g = 1$ to simplify the presentation. This expression can be inverted to yield
\begin{equation}
  \label{eq:14}
  r_* \ = \ \sqrt{\frac{4a^2}{(1 + a)(1 - 3a)}}\;,
\end{equation}
with the positive branch being selected on physical grounds. Reality of the black-hole horizon sets a bound on $a$
\begin{align}\label{bound}
  -1 < a < \frac{1}{3}
\end{align}
and since this is a radius we can see that $a$ defines the range of $r_*$ to be $[0,\infty)$.

At this stage we should also note that in order to preserve the signature of the metric \eqref{BHmetric}, we have to restrict $\Xi$ such that it is always larger than zero. This implies that
\begin{align}
  \Xi \ &= \ 1 - a^2 > 0 \;,
\end{align}
which is always satisfied for a choice in \eqref{bound}.

Returning to \eqref{discriminant}, one finds that it can be written as
\begin{align}
	C_1^2 - 4 C_0 C_2 \ =& \ -\frac{1024  r_*^4 \left(r_+^2-r_*^2\right){}^2 \left( r_*^2+1\right) \left(4 r_+^2 \left(- \sqrt{ r_*^2 \left(
   r_*^2+1\right)}+ r_*^2+1\right)-r_*^2\right)}{9 \left( r_*^2 - 2 \sqrt{r_*^2 \left( r_*^2+1\right)}\right){}^4 \left(
    4\sqrt{ r_*^2 \left( r_*^2+1\right)}+ r_*^2\right){}^2} \, .
\end{align}
All of the above terms are clearly positive, with the exception of 
\begin{align} \label{two}
  \lambda \ \equiv \ 4 r_+^2 \left(- \sqrt{ r_*^2 \left(r_*^2+1\right)}+ r_*^2+1\right)-r_*^2 \, .
\end{align}
Let us take a part of this and define
\begin{align}
  f(r_*) \ = \ 4(- \sqrt{r_*^2\left(r_*^2+1\right)}+r_*^2 +1) \, .
\end{align}
The range for this function is $2 < f(r_*) <4$. This can be seen by taking the infinite limit, or by Laurent expansion. Hence,
\begin{align}
  \lambda > 2 r_+^2-r_*^2 > 0 \, ,
\end{align}
and the whole of \eqref{discriminant} is always negative except at the BPS limit. We have therefore shown that the mass is indeed complex for $r_+ < r_*$.

\section{5-sphere Geometry}\label{AppB}

In this appendix we give the details of the change of coordinates \eqref{usingApp}. Start from $\mathbb{C}^3$ with complex coordinates $(z_1,z_2,z_3)$ and metric
\begin{align}
	\dd s^2_6 \ = \ \sum_{i=1}^3 | \dd z_i |^2 \, .
\end{align}
We then restrict to $S^5$ by imposing the constraint $\sum_{i=1}^3 | z_i |^2 =1$. Next define coordinates
\begin{align}
	\xi_1 \ = \ \frac{z_1}{z_3} \, , \qquad \xi_2 \ = \ \frac{z_2}{z_3} \, , \qquad z_3 \ = \ | z_3 | e^{i \tau} \, ,
\end{align}
and 
\begin{align}
	\hat{A} \ =& \ \frac{i}{2} ( 1 + | \xi_1 |^2 + | \xi_2 |^2 )^{-1} \left( \bar{\xi}_1 \dd \xi_1 + \bar{\xi}_2 \dd \xi_2 - \xi_1 \dd \bar{\xi}_1 - \xi_2 \dd \bar{\xi_2} \right) \, .
\end{align}
Then the 5-sphere metric is
\begin{align}\begin{split}
	\dd s^2_{S^5} \ =& \ ( \dd \tau - \hat{A} )^2 + ( 1 + | \xi_1 |^2 + | \xi_2 |^2 )^{-1} ( | \dd \xi_1 |^2 + | \dd \xi_2 |^2 )  \\
	&\ - ( 1 + | \xi_1 |^2 + | \xi_2 |^2 )^{-2} ( \bar{\xi}_1 \dd \xi_1 + \bar{\xi}_2 \dd \xi_2 )^2\;.
\end{split}\end{align}
Choosing
\begin{align}\begin{split}
	\xi_1 \ =& \ \tan \chi \cos \frac{\theta}{2} e^{\frac{i}{2}(\psi+\varphi)}  \\
	\xi_2 \ =& \ \tan \chi \sin \frac{\theta}{2} e^{\frac{i}{2}(\psi-\varphi)} \, ,
\end{split}\end{align}
gives 
\begin{align}\begin{split}
	\dd s^2_{S^5} \ =& \ \left( \dd \tau + \frac{1}{2} \sin^2 \chi ( \dd \psi + \cos \theta \dd \varphi) \right)^2 \\
	&\ + \dd \chi^2 + \frac{1}{4} \sin^2 \chi \Big[ \cos^2 \chi ( \dd \psi + \cos \theta \dd \varphi )^2 + \dd \theta^2 + \sin^2 \theta \dd \varphi^2 \Big] \, , \\
	=& \ \dd \tau^2 + \sin^2 \chi \dd \tau ( \dd \psi + \cos \theta \dd \varphi) \\
	&\ + \dd \chi^2 + \frac{1}{4} \sin^2 \chi \Big[ ( \dd \psi + \cos \theta \dd \varphi )^2 + \dd \theta^2 + \sin^2 \theta \dd \varphi^2 \Big] \, . \label{S5metric}
\end{split}\end{align}

From the choice of $\xi$'s we get 
\begin{align}\begin{split}
	z_1 \ =& \ \sin \chi \cos \frac{\theta}{2} e^{\frac{i}{2}(\psi+\varphi)} e^{i \tau}  \\
	z_2 \ =& \ \sin \chi \sin \frac{\theta}{2} e^{\frac{i}{2}(\psi-\varphi)} e^{i \tau}  \\
	z_3 \ =& \ \cos \chi e^{i \tau} \, .
\end{split}\end{align}

On the other hand one can take the following Cartesian coordinates for $S^5$
\begin{align}\begin{split}
	x_1 \ =& \ \sin \phi_1 \sin \theta_1 \sin \theta_2  \\
	x_2 \ =& \ \cos \phi_1 \sin \theta_1 \sin \theta_2  \\
	x_3 \ =& \ \sin \phi_2 \sin \theta_1 \cos \theta_2  \\
	x_4 \ =& \ \cos \phi_2 \sin \theta_1 \cos \theta_2 \\
	x_5 \ =& \ \sin \phi_3 \cos \theta_1  \\
	x_6 \ =& \ \cos \phi_3 \cos \theta_1 \, ,
\end{split}\end{align}
and identify
\begin{align}\begin{split}
	\tilde{z}_1 \ =& \ x_2 + i x_1 \ = \ e^{i \phi_1} \sin \theta_1 \sin \theta_2  \\
	\tilde{z}_2 \ =& \ x_4 + i x_3 \ = \ e^{i \phi_2} \sin \theta_1 \cos \theta_2  \\
	\tilde{z}_3 \ =& \ x_6 + i x_5 \ = \ e^{i \phi_3} \cos \theta_1 \, .
\end{split}\end{align}
The metric in these coordinates is 
\begin{equation}
	\dd s_{S^5}^2 \ = \ \dd \theta_1^2 + \sin^2 \theta_1 \dd \theta_2^2 + \sin^2 \theta_1 \sin^2 \theta_2 \dd \phi_1^2 + \sin^2 \theta_1 \cos^2 \theta_2 \dd \phi_2^2 + \cos^2 \theta_1 \dd \phi_3^2 \, .  \label{S5metric2}
\end{equation}
Inserting the coordinate mapping
\begin{align}
	\theta \ \equiv \ 2 \theta_2 \, , \quad \psi \ \equiv \ \phi_1 + \phi_2 - 2 \phi_3 \, , \quad \varphi \ \equiv \ - \phi_1 + \phi_2 \, , \quad \tau \ \equiv \ \phi_3 \, , \quad \chi \ \equiv \ \theta_1 \, .
\end{align}
into \eqref{S5metric} gives back \eqref{S5metric2}.

\section{5D 1-loop Determinants for General $n_0$}\label{AppC}

In this appendix we provide some additional details on the derivation of the 1-loop determinants of 5D MSYM on a squashed $S^5$ for general $n_0$ and their simplification in the Cardy-like limit, $\omega\to 0$, which we made use of in Sec.~\ref{5Dderiv}. We will follow both \cite{Imamura:2012bm} and \cite{Bobev:2015kza} closely, to which we refer the interested reader for a complete account. The calculation of \cite{Imamura:2012bm} involves looking at the partition function of a 6D (1,0) theory with vector and hypermultiplets on $S^5 \times S^1$, calculating the 6D one-loop determinants via supersymmetric localisation and then reducing on the circle to get expressions in 5D. Interestingly, even though the 6D theory is not superconformal, the result for the 1-loop determinants can be expressed as a Hamiltonian index. This fact greatly simplifies the organisation of the calculation for $n_0=0$ and allows for a straightforward extension to arbitrary $n_0$.

We begin by setting up notation, as in \cite{Imamura:2012bm}. There are 5 Cartans left after a choice of localising supercharge $\mathcal{Q}$ in the 6D (1,0) theory
\begin{align}
  H = -\partial_t, \quad Q_V = -i \mathcal{L}_{\psi}, \quad \tau_3, \quad \lambda_{3,8}\;.
\end{align}
These Cartans correspond to the bosonic symmetries $\mathbb{R}\times U(1)_V \times SU(3)_V \times SU(2)_R$.  We will also make the supercharge we picked manifest, with the associated supersymmetry parameters being
\begin{center}
\begin{tabular}{ |c|c|c|c|c|c| }
 \hline
  & $H$ & $Q_V$ & $\tau_3$ & $\lambda_3$ & $\lambda_8$ \\ [0.5ex]
 \hline
 $\varepsilon_1$ & $+\frac{1}{2}$ & $-\frac{3}{2}$ & +1 & 0 & 0 \\ [0.5ex]
 $\varepsilon_2$ & $-\frac{1}{2}$ & $+\frac{3}{2}$ & -1 & 0 & 0 \\ [0.5ex]
 \hline
\end{tabular}
\end{center}
With this information, one can formulate the following ``index''
\cite{Imamura:2012bm}\footnote{When the theory is superconformal this
  is the standard 6D (1,0) index.}
\begin{align} \label{eq:c2}
  \mathcal{I} = \Tr (-1)^F q^{H-Q_V-2\tau_3}x^{Q_V + \frac{3}{2}\tau_3}y_3^{\lambda_3}y_8^{\lambda_8}\;,
\end{align}
where $q=e^{-\beta}$, $x = q^{1+i w_0}$ and $y_{3,8} = q^{i w_{3,8}}$. The trace is taken over the Fock space of gauge-invariant states on the squashed  $S^5$. We choose the fermion-number operator as $F = 2Q_V$. This index uses completely independent chemical potentials since the operator on the RHS of \eqref{eq:c2}  commutes with the supercharge $\QQ$. We note that the chemical potentials $w_{0,3,8}$ are related to the $\omega_{1,2,3}$ that appear in the main part of this paper as
\begin{align}
  \label{eq:34}
  (1+iw_0)=\frac{1}{3}\sum_{i = 1}^3\omega_i\;,\qquad  i w_3 =\frac{1}{2}(\omega_2 - \omega_1)\;,\qquad i w_8 = \frac{1}{6}(2 \omega_3 - \omega_1 - \omega_2)\;
\end{align}
and these are in turn related to the squashing parameters $\varphi_i$ as $\omega_i = 1 + i \varphi_i$.

Following the discussion in Sec.~\ref{indexcft}, one can straightforwardly generalise \eqref{eq:c2} for arbitrary $n_0$ by shifting one of the chemical potentials, $\omega_1 \to \omega_1 - \frac{2 \pi i n_0}{\beta} =  \tilde{\omega}_1$
\begin{align} \label{eq:c1}
  \mathcal{I} = \Tr (-1)^F q^{H-Q_V-2\tau_3}\tilde{x}^{Q_V + \frac{3}{2}\tau_3}\tilde{y}_3^{\lambda_3} \tilde{y}_8^{\lambda_8}\;,
\end{align}
with the fugacities $\tilde{x}, \tilde{y}_{3,8}$  defined as below \eqref{eq:c2} but using the $\tilde{\omega}_i$ of \eqref{eq:6}. From here on we can repeat the calculation of \cite{Imamura:2012bm} using the tilded chemical potentials. In particular, this means that we will encounter the same bosonic and fermionic cancellations that occur in the calculation of the supersymmetric partition function of \cite{Imamura:2012bm}. The non-cancelling  contributions from the vector multiplet are fermionic modes that can be encoded into the plethystic exponential of the single-letter index
\begin{align}
  &-\sum_{\lambda \in \text{adj}}\left[\sum_{k=1}^{\infty}q^{-i\lambda(\sigma)}\tilde{x}^{k}\chi_{(k,k)}(\tilde{y}_3,\tilde{y}_8) + \sum_{k=0}^{\infty}q^{-i\lambda(\sigma)}\tilde{x}^{(k + 3)}\chi_{(k,k)}(\tilde{y}_3,\tilde{y}_8)\right]\cr
  & = -\sum_{\lambda \in \text{adj}}\left[\sum_{k=1}^{\infty}q^{-i\lambda(\sigma)}\tilde{x}^{k}\chi_{(k,k)}(\tilde{y}_3,\tilde{y}_8) + \sum_{k=0}^{\infty}q^{-i\lambda(\sigma) + \sum_i \omega_i}\tilde{x}^{k}\chi_{(k,k)}(\tilde{y}_3,\tilde{y}_8)\right]\;,
\end{align}
where the $\chi_{(k,k)}(\tilde{y}_3,\tilde{y}_8)$ is the $SU(3)_V$ character and the $\lambda(\sigma)$ denote an inner product in weight space. The surviving modes from the hypermultiplet in a representation $R$ have a corresponding single-letter index contribution
\begin{align}
  \sum_{\rho \in R}\left[\sum_{k=0}^{\infty}q^{-i\rho(\sigma)}\tilde{x}^{(k + \frac{3}{2})}\chi_{(k,k)}(\tilde{y}_3,\tilde{y}_8) + \sum_{k=0}^{\infty}q^{i\rho(\sigma)}\tilde{x}^{(k + \frac{3}{2})}\chi_{(k,k)}(\tilde{y}_3,\tilde{y}_8)\right]\;.
\end{align}
We can use the fact that $n_0 \in \mathbb{Z}$ and recast this as 
\begin{align}
  \sum_{\rho \in R}\left[\sum_{k=0}^{\infty}q^{-i\rho(\sigma)+ \frac{1}{2}\sum_i \omega_i -\frac{\pi i n_0}{\beta}}\tilde{x}^{k }\chi_{(k,k)}(\tilde{y}_3,\tilde{y}_8) + \sum_{k=0}^{\infty}q^{i\rho(\sigma)+ \frac{1}{2}\sum_i \omega_i+\frac{\pi i n_0}{\beta}}\tilde{x}^{k }\chi_{(k,k)}(\tilde{y}_3,\tilde{y}_8)\right]\;.
\end{align}

By grouping the $q^{\pm (i\rho(\sigma)-\frac{\pi i n_0}{\beta})}$ expressions together in the hypermultiplet, the calculation of the corresponding 1-loop determinants in \cite{Imamura:2012bm} can be carried out identically. The only point that  merits special mention is that the $p_i$ variables of Eq.~(79) of that reference are still the ones that are used in the general $n_0$ case, since
\begin{align}
  \label{eq:34}
  \tilde{p}_i = q^{\tilde{\omega}_i} = q^{\omega_i}\;.
\end{align}
Following the rest of the arguments in \cite{Imamura:2012bm}, one ends up with the final result for the 1-loop determinants of a 5D $\mathcal N=1$ theory with one vector and one hypermultiplet in the representation $R$
\begin{align}
  \mathcal Z_{1-\text{loop}} = \frac{\prod_{\alpha \in \text{roots}} S_3(-i\lambda(\sigma)|\vec\omega)}{\prod_{\rho \in \text{R}} S_3(-i\rho(\sigma) - \frac{i \pi n_0}{\beta} + \frac{\omega_1+\omega_2+\omega_3}{2}|\vec\omega)}\;.
\end{align}

In the Cardy-like limit $\omega_i \to 0$, the triple-sine
functions\footnote{For a summary of the properties of triple-sine
  functions see also \cite{Tizzano:2014roa}.} $S_3(\lambda|\vec \omega)$ have a simple  $\beta\to \infty $ limit if one additionally assumes that the $\lambda$ scale as $\beta$, while the remaining parameters remain of order one \cite{Bobev:2015kza}
\begin{align}
  \log S_3( i \lambda|\vec{\omega}) \overset{sgn(\lambda)= \pm 1}{\sim}-\frac{\pi}{6\omega_1 \omega_2 \omega_3} \left(|\lambda|^3 +O(\beta^2, \vec\omega)\right) \;.
\end{align}
As a result Eq.~\eqref{eq:31} from the main part of this paper where $\vec \omega = (\omega, \omega, \omega)$ becomes
\begin{align}
  \mathcal Z^{n_0}_{\text{1-loop}} &\propto \prod_{i>j}^N \frac{S_3(i \lambda_{ij}|\vec{\omega})S_3(-i\lambda_{ij}|\vec{\omega})}{S_3(\phi+i\lambda_{ij}|\vec{\omega}) S_3(\phi-i\lambda_{ij}|\vec{\omega})}  \cr
  &\sim \exp \left( -\frac{\pi}{6\omega^3}\sum_{i>j}\left[2 \lambda_{ij}^3 - (\lambda_{ij} - i\phi)^3 - (\lambda_{ij}+ i\phi)^3\right] + O(\beta,\omega)\right)\cr
  &= \exp \left(-\frac{\pi}{\omega^3}\phi^2 \sum_{i>j}\lambda_{ij} +O(\beta, \omega)\right)\;,
\end{align}
where the absolute values have been dropped because the parameters $\lambda$ have been restricted to the Weyl chamber where $\lambda_i >\lambda_j$ for $i>j$ and since we have assumed that $\lambda_{ij}\gg  i \phi$ because of its $\beta$ scaling. Note that in the Cardy-like limit $\omega\to 0$ the constraint implies that $\phi$ is imaginary, so the combination $\lambda_{ij} \pm i \phi $ is real and positive.

\end{appendix}


\newpage

\bibliography{6Dindex_v7}

\begin{thebibliography}{10}
\ifx\href\asklfhas\newcommand{\href}[2]{#2}\fi
\ifx\arxivref\asklfhas\newcommand{\arxivref}[2]{\href{http://arxiv.org/abs/#1}{#2}}\fi
\ifx\doiref\asklfhas\newcommand{\doiref}[2]{\href{http://dx.doi.org/#1}{#2}}\fi
\parskip 0pt
\normalsize

\bibitem{Kinney:2005ej}
J.~Kinney, J.~M. Maldacena, S.~Minwalla \& S.~Raju,
\textit{``{An Index for 4 dimensional super conformal theories}''},
\doiref{10.1007/s00220-007-0258-7}{Commun.~Math.~Phys. \textbf{275}, 209
  (2007)\ignorespaces}\ignorespaces,
\normalsize{\texttt{\arxivref{hep-th/0510251}{hep-th/0510251}}}\ignorespaces
\bibitem{Romelsberger:2005eg}
C.~Romelsberger,
\textit{``{Counting chiral primaries in N = 1, d=4 superconformal field
  theories}''},
\doiref{10.1016/j.nuclphysb.2006.03.037}{Nucl.~Phys. \textbf{B747}, 329
  (2006)\ignorespaces}\ignorespaces,
\normalsize{\texttt{\arxivref{hep-th/0510060}{hep-th/0510060}}}\ignorespaces
\bibitem{Bhattacharya:2008zy}
J.~Bhattacharya, S.~Bhattacharyya, S.~Minwalla \& S.~Raju,
\textit{``{Indices for Superconformal Field Theories in 3,5 and 6
  Dimensions}''},
\doiref{10.1088/1126-6708/2008/02/064}{JHEP \textbf{0802}, 064
  (2008)\ignorespaces}\ignorespaces,
\normalsize{\texttt{\arxivref{0801.1435}{arXiv:0801.1435}}}\ignorespaces
\bibitem{Grant:2008sk}
L.~Grant, P.~A. Grassi, S.~Kim \& S.~Minwalla,
\textit{``{Comments on 1/16 BPS Quantum States and Classical
  Configurations}''},
\doiref{10.1088/1126-6708/2008/05/049}{JHEP \textbf{0805}, 049
  (2008)\ignorespaces}\ignorespaces,
\normalsize{\texttt{\arxivref{0803.4183}{arXiv:0803.4183}}}\ignorespaces
\bibitem{Chang:2013fba}
C.-M. Chang \& X.~Yin,
\textit{``{1/16 BPS states in $\mathcal N=$ 4 super-Yang-Mills theory}''},
\doiref{10.1103/PhysRevD.88.106005}{Phys.~Rev. \textbf{D88}, 106005
  (2013)\ignorespaces}\ignorespaces,
\normalsize{\texttt{\arxivref{1305.6314}{arXiv:1305.6314}}}\ignorespaces
\bibitem{Nawata:2011un}
S.~Nawata,
\textit{``{Localization of N=4 Superconformal Field Theory on $S^1 x S^3$ and
  Index}''},
\doiref{10.1007/JHEP11(2011)144}{JHEP \textbf{1111}, 144
  (2011)\ignorespaces}\ignorespaces,
\normalsize{\texttt{\arxivref{1104.4470}{arXiv:1104.4470}}}\ignorespaces
\bibitem{Kim:2012ava}
H.-C. Kim \& S.~Kim,
\textit{``{M5-branes from gauge theories on the 5-sphere}''},
\doiref{10.1007/JHEP05(2013)144}{JHEP \textbf{1305}, 144
  (2013)\ignorespaces}\ignorespaces,
\normalsize{\texttt{\arxivref{1206.6339}{arXiv:1206.6339}}}\ignorespaces
\bibitem{Lorenzen:2014pna}
J.~Lorenzen \& D.~Martelli,
\textit{``{Comments on the Casimir energy in supersymmetric field theories}''},
\doiref{10.1007/JHEP07(2015)001}{JHEP \textbf{1507}, 001
  (2015)\ignorespaces}\ignorespaces,
\normalsize{\texttt{\arxivref{1412.7463}{arXiv:1412.7463}}}\ignorespaces
\bibitem{Assel:2015nca}
B.~Assel, D.~Cassani, L.~Di~Pietro, Z.~Komargodski, J.~Lorenzen \& D.~Martelli,
\textit{``{The Casimir Energy in Curved Space and its Supersymmetric
  Counterpart}''},
\doiref{10.1007/JHEP07(2015)043}{JHEP \textbf{1507}, 043
  (2015)\ignorespaces}\ignorespaces,
\normalsize{\texttt{\arxivref{1503.05537}{arXiv:1503.05537}}}\ignorespaces
\bibitem{Cassani:2014zwa}
D.~Cassani \& D.~Martelli,
\textit{``{The gravity dual of supersymmetric gauge theories on a squashed
  S$^{1}$ x S$^{3}$}''},
\doiref{10.1007/JHEP08(2014)044}{JHEP \textbf{1408}, 044
  (2014)\ignorespaces}\ignorespaces,
\normalsize{\texttt{\arxivref{1402.2278}{arXiv:1402.2278}}}\ignorespaces
\bibitem{Hosseini:2017mds}
S.~M. Hosseini, K.~Hristov \& A.~Zaffaroni,
\textit{``{An extremization principle for the entropy of rotating BPS black
  holes in AdS$_{5}$}''},
\doiref{10.1007/JHEP07(2017)106}{JHEP \textbf{1707}, 106
  (2017)\ignorespaces}\ignorespaces,
\normalsize{\texttt{\arxivref{1705.05383}{arXiv:1705.05383}}}\ignorespaces
\bibitem{Hosseini:2018dob}
S.~M. Hosseini, K.~Hristov \& A.~Zaffaroni,
\textit{``{A note on the entropy of rotating BPS AdS$_7\times S^4$ black
  holes}''},
\doiref{10.1007/JHEP05(2018)121}{JHEP \textbf{1805}, 121
  (2018)\ignorespaces}\ignorespaces,
\normalsize{\texttt{\arxivref{1803.07568}{arXiv:1803.07568}}}\ignorespaces
\bibitem{Cabo-Bizet:2018ehj}
A.~Cabo-Bizet, D.~Cassani, D.~Martelli \& S.~Murthy,
\textit{``{Microscopic origin of the Bekenstein-Hawking entropy of
  supersymmetric AdS$_{\bf 5}$ black holes}''},
\normalsize{\texttt{\arxivref{1810.11442}{arXiv:1810.11442}}}\ignorespaces
\bibitem{Cassani:2019mms}
D.~Cassani \& L.~Papini,
\textit{``{The BPS limit of rotating AdS black hole thermodynamics}''},
\normalsize{\texttt{\arxivref{1906.10148}{arXiv:1906.10148}}}\ignorespaces
\bibitem{Larsen:2019oll}
F.~Larsen, J.~Nian \& Y.~Zeng,
\textit{``{AdS$_5$ Black Hole Entropy near the BPS Limit}''},
\normalsize{\texttt{\arxivref{1907.02505}{arXiv:1907.02505}}}\ignorespaces
\bibitem{Choi:2018hmj}
S.~Choi, J.~Kim, S.~Kim \& J.~Nahmgoong,
\textit{``{Large AdS black holes from QFT}''},
\normalsize{\texttt{\arxivref{1810.12067}{arXiv:1810.12067}}}\ignorespaces
\bibitem{Choi:2018fdc}
S.~Choi, C.~Hwang, S.~Kim \& J.~Nahmgoong,
\textit{``{Entropy functions of BPS black holes in AdS$_4$ and AdS$_6$}''},
\normalsize{\texttt{\arxivref{1811.02158}{arXiv:1811.02158}}}\ignorespaces
\bibitem{Benini:2018mlo}
F.~Benini \& P.~Milan,
\textit{``{A Bethe Ansatz type formula for the superconformal index}''},
\normalsize{\texttt{\arxivref{1811.04107}{arXiv:1811.04107}}}\ignorespaces
\bibitem{Benini:2018ywd}
F.~Benini \& P.~Milan,
\textit{``{Black holes in 4d $\mathcal{N}=4$ Super-Yang-Mills}''},
\normalsize{\texttt{\arxivref{1812.09613}{arXiv:1812.09613}}}\ignorespaces
\bibitem{Kim:2019yrz}
J.~Kim, S.~Kim \& J.~Song,
\textit{``{A 4d $N=1$ Cardy Formula}''},
\normalsize{\texttt{\arxivref{1904.03455}{arXiv:1904.03455}}}\ignorespaces
\bibitem{Cabo-Bizet:2019osg}
A.~Cabo-Bizet, D.~Cassani, D.~Martelli \& S.~Murthy,
\textit{``{The asymptotic growth of states of the 4d N=1 superconformal
  index}''},
\normalsize{\texttt{\arxivref{1904.05865}{arXiv:1904.05865}}}\ignorespaces
\bibitem{Lockhart:2012vp}
G.~Lockhart \& C.~Vafa,
\textit{``{Superconformal Partition Functions and Non-perturbative Topological
  Strings}''},
\normalsize{\texttt{\arxivref{1210.5909}{arXiv:1210.5909}}}\ignorespaces
\bibitem{Kallen:2012cs}
J.~Källén \& M.~Zabzine,
\textit{``{Twisted supersymmetric 5D Yang-Mills theory and contact
  geometry}''},
\doiref{10.1007/JHEP05(2012)125}{JHEP \textbf{1205}, 125
  (2012)\ignorespaces}\ignorespaces,
\normalsize{\texttt{\arxivref{1202.1956}{arXiv:1202.1956}}}\ignorespaces
\bibitem{Kallen:2012va}
J.~Källén, J.~Qiu \& M.~Zabzine,
\textit{``{The perturbative partition function of supersymmetric 5D Yang-Mills
  theory with matter on the five-sphere}''},
\doiref{10.1007/JHEP08(2012)157}{JHEP \textbf{1208}, 157
  (2012)\ignorespaces}\ignorespaces,
\normalsize{\texttt{\arxivref{1206.6008}{arXiv:1206.6008}}}\ignorespaces
\bibitem{Kallen:2012zn}
J.~Källén, J.~A. Minahan, A.~Nedelin \& M.~Zabzine,
\textit{``{$N^3$-behavior from 5D Yang-Mills theory}''},
\doiref{10.1007/JHEP10(2012)184}{JHEP \textbf{1210}, 184
  (2012)\ignorespaces}\ignorespaces,
\normalsize{\texttt{\arxivref{1207.3763}{arXiv:1207.3763}}}\ignorespaces
\bibitem{Kim:2012qf}
H.-C. Kim, J.~Kim \& S.~Kim,
\textit{``{Instantons on the 5-sphere and M5-branes}''},
\normalsize{\texttt{\arxivref{1211.0144}{arXiv:1211.0144}}}\ignorespaces
\bibitem{Kim:2013nva}
H.-C. Kim, S.~Kim, S.-S. Kim \& K.~Lee,
\textit{``{The general M5-brane superconformal index}''},
\normalsize{\texttt{\arxivref{1307.7660}{arXiv:1307.7660}}}\ignorespaces
\bibitem{Bobev:2015kza}
N.~Bobev, M.~Bullimore \& H.-C. Kim,
\textit{``{Supersymmetric Casimir Energy and the Anomaly Polynomial}''},
\doiref{10.1007/JHEP09(2015)142}{JHEP \textbf{1509}, 142
  (2015)\ignorespaces}\ignorespaces,
\normalsize{\texttt{\arxivref{1507.08553}{arXiv:1507.08553}}}\ignorespaces
\bibitem{Jafferis:2012iv}
D.~L. Jafferis \& S.~S. Pufu,
\textit{``{Exact results for five-dimensional superconformal field theories
  with gravity duals}''},
\doiref{10.1007/JHEP05(2014)032}{JHEP \textbf{1405}, 032
  (2014)\ignorespaces}\ignorespaces,
\normalsize{\texttt{\arxivref{1207.4359}{arXiv:1207.4359}}}\ignorespaces
\bibitem{Chang:2014kma}
C.-M. Chang, Y.-H. Lin, Y.~Wang \& X.~Yin,
\textit{``{Deformations with Maximal Supersymmetries Part 1: On-shell
  Formulation}''},
\normalsize{\texttt{\arxivref{1403.0545}{arXiv:1403.0545}}}\ignorespaces
\bibitem{Lin:2015zea}
Y.-H. Lin, S.-H. Shao, Y.~Wang \& X.~Yin,
\textit{``{Interpolating the Coulomb Phase of Little String Theory}''},
\doiref{10.1007/JHEP12(2015)022}{JHEP \textbf{1512}, 022
  (2015)\ignorespaces}\ignorespaces,
\normalsize{\texttt{\arxivref{1502.01751}{arXiv:1502.01751}}}\ignorespaces
\bibitem{Mezei:2018url}
M.~Mezei, S.~S. Pufu \& Y.~Wang,
\textit{``{Chern-Simons theory from M5-branes and calibrated M2-branes}''},
\normalsize{\texttt{\arxivref{1812.07572}{arXiv:1812.07572}}}\ignorespaces
\bibitem{Douglas:2010iu}
M.~R. Douglas,
\textit{``{On D=5 super Yang-Mills theory and (2,0) theory}''},
\doiref{10.1007/JHEP02(2011)011}{JHEP \textbf{1102}, 011
  (2011)\ignorespaces}\ignorespaces,
\normalsize{\texttt{\arxivref{1012.2880}{arXiv:1012.2880}}}\ignorespaces
\bibitem{Lambert:2010iw}
N.~Lambert, C.~Papageorgakis \& M.~Schmidt-Sommerfeld,
\textit{``{M5-Branes, D4-Branes and Quantum 5D super-Yang-Mills}''},
\doiref{10.1007/JHEP01(2011)083}{JHEP \textbf{1101}, 083
  (2011)\ignorespaces}\ignorespaces,
\normalsize{\texttt{\arxivref{1012.2882}{arXiv:1012.2882}}}\ignorespaces
\bibitem{Papageorgakis:2014dma}
C.~Papageorgakis \& A.~B. Royston,
\textit{``{Revisiting Soliton Contributions to Perturbative Amplitudes}''},
\doiref{10.1007/JHEP09(2014)128}{JHEP \textbf{1409}, 128
  (2014)\ignorespaces}\ignorespaces,
\normalsize{\texttt{\arxivref{1404.0016}{arXiv:1404.0016}}}\ignorespaces
\bibitem{Sen:2007qy}
A.~Sen,
\textit{``{Black Hole Entropy Function, Attractors and Precision Counting of
  Microstates}''},
\doiref{10.1007/s10714-008-0626-4}{Gen.~Rel.~Grav. \textbf{40}, 2249
  (2008)\ignorespaces}\ignorespaces,
\normalsize{\texttt{\arxivref{0708.1270}{arXiv:0708.1270}}}\ignorespaces
\bibitem{Sen:2008yk}
A.~Sen,
\textit{``{Entropy Function and AdS(2) / CFT(1) Correspondence}''},
\doiref{10.1088/1126-6708/2008/11/075}{JHEP \textbf{0811}, 075
  (2008)\ignorespaces}\ignorespaces,
\normalsize{\texttt{\arxivref{0805.0095}{arXiv:0805.0095}}}\ignorespaces
\bibitem{Benini:2015noa}
F.~Benini \& A.~Zaffaroni,
\textit{``{A topologically twisted index for three-dimensional supersymmetric
  theories}''},
\doiref{10.1007/JHEP07(2015)127}{JHEP \textbf{1507}, 127
  (2015)\ignorespaces}\ignorespaces,
\normalsize{\texttt{\arxivref{1504.03698}{arXiv:1504.03698}}}\ignorespaces
\bibitem{Benini:2015eyy}
F.~Benini, K.~Hristov \& A.~Zaffaroni,
\textit{``{Black hole microstates in AdS$_{4}$ from supersymmetric
  localization}''},
\doiref{10.1007/JHEP05(2016)054}{JHEP \textbf{1605}, 054
  (2016)\ignorespaces}\ignorespaces,
\normalsize{\texttt{\arxivref{1511.04085}{arXiv:1511.04085}}}\ignorespaces
\bibitem{Hosseini:2018uzp}
S.~M. Hosseini, I.~Yaakov \& A.~Zaffaroni,
\textit{``{Topologically twisted indices in five dimensions and holography}''},
\doiref{10.1007/JHEP11(2018)119}{JHEP \textbf{1811}, 119
  (2018)\ignorespaces}\ignorespaces,
\normalsize{\texttt{\arxivref{1808.06626}{arXiv:1808.06626}}}\ignorespaces
\bibitem{Townsend:1983kk}
P.~K. Townsend \& P.~van~Nieuwenhuizen,
\textit{``{Gauged Seven-Dimensional Supergravity}''},
\doiref{10.1016/0370-2693(83)91230-3}{Phys.~Lett. \textbf{B125}, 41
  (1983)\ignorespaces}\ignorespaces
\bibitem{Chong:2004dy}
Z.~W. Chong, M.~Cvetic, H.~Lu \& C.~N. Pope,
\textit{``{Non-extremal charged rotating black holes in seven-dimensional
  gauged supergravity}''},
\doiref{10.1016/j.physletb.2005.07.054}{Phys.~Lett. \textbf{B626}, 215
  (2005)\ignorespaces}\ignorespaces,
\normalsize{\texttt{\arxivref{hep-th/0412094}{hep-th/0412094}}}\ignorespaces
\bibitem{Chow:2007ts}
D.~D.~K. Chow,
\textit{``{Equal charge black holes and seven dimensional gauged
  supergravity}''},
\doiref{10.1088/0264-9381/25/17/175010}{Class.~Quant.~Grav. \textbf{25}, 175010
  (2008)\ignorespaces}\ignorespaces,
\normalsize{\texttt{\arxivref{0711.1975}{arXiv:0711.1975}}}\ignorespaces
\bibitem{Cvetic:2005zi}
M.~Cvetic, G.~W. Gibbons, H.~Lu \& C.~N. Pope,
\textit{``{Rotating black holes in gauged supergravities: Thermodynamics,
  supersymmetric limits, topological solitons and time machines}''},
\normalsize{\texttt{\arxivref{hep-th/0504080}{hep-th/0504080}}}\ignorespaces
\bibitem{Chen:2005zj}
W.~Chen, H.~Lu \& C.~N. Pope,
\textit{``{Mass of rotating black holes in gauged supergravities}''},
\doiref{10.1103/PhysRevD.73.104036}{Phys.~Rev. \textbf{D73}, 104036
  (2006)\ignorespaces}\ignorespaces,
\normalsize{\texttt{\arxivref{hep-th/0510081}{hep-th/0510081}}}\ignorespaces
\bibitem{Liu:1999ai}
J.~T. Liu \& R.~Minasian,
\textit{``{Black holes and membranes in AdS(7)}''},
\doiref{10.1016/S0370-2693(99)00500-6}{Phys.~Lett. \textbf{B457}, 39
  (1999)\ignorespaces}\ignorespaces,
\normalsize{\texttt{\arxivref{hep-th/9903269}{hep-th/9903269}}}\ignorespaces
\bibitem{Witten:1998qj}
E.~Witten,
\textit{``{Anti-de Sitter space and holography}''},
\doiref{10.4310/ATMP.1998.v2.n2.a2}{Adv.~Theor.~Math.~Phys. \textbf{2}, 253
  (1998)\ignorespaces}\ignorespaces,
\normalsize{\texttt{\arxivref{hep-th/9802150}{hep-th/9802150}}}\ignorespaces
\bibitem{Imamura:2012bm}
Y.~Imamura,
\textit{``{Perturbative partition function for squashed $S^5$}''},
\doiref{10.1093/ptep/ptt044}{PTEP \textbf{2013}, 073B01
  (2013)\ignorespaces}\ignorespaces,
\normalsize{\texttt{\arxivref{1210.6308}{arXiv:1210.6308}}}\ignorespaces
\bibitem{Bullimore:2014upa}
M.~Bullimore \& H.-C. Kim,
\textit{``{The Superconformal Index of the (2,0) Theory with Defects}''},
\doiref{10.1007/JHEP05(2015)048}{JHEP \textbf{1505}, 048
  (2015)\ignorespaces}\ignorespaces,
\normalsize{\texttt{\arxivref{1412.3872}{arXiv:1412.3872}}}\ignorespaces
\bibitem{Beem:2014kka}
C.~Beem, L.~Rastelli \& B.~C. van~Rees,
\textit{``{$ \mathcal{W} $ symmetry in six dimensions}''},
\doiref{10.1007/JHEP05(2015)017}{JHEP \textbf{1505}, 017
  (2015)\ignorespaces}\ignorespaces,
\normalsize{\texttt{\arxivref{1404.1079}{arXiv:1404.1079}}}\ignorespaces
\bibitem{Kim:2016usy}
S.~Kim \& K.~Lee,
\textit{``{Indices for 6 dimensional superconformal field theories}''},
\normalsize{\texttt{\arxivref{1608.02969}{arXiv:1608.02969}}}\ignorespaces,
in \textit{``{Localization Techniques in Quantum Field Theories}''}
\bibitem{Pestun:2016zxk}
V.~Pestun et~al.,
\textit{``{Localization Techniques in Quantum Field Theories}''},
\normalsize{\texttt{\arxivref{1608.02952}{arXiv:1608.02952}}}\ignorespaces
\bibitem{Alday:2014rxa}
L.~F. Alday, M.~Fluder, P.~Richmond \& J.~Sparks,
\textit{``{Gravity Dual of Supersymmetric Gauge Theories on a Squashed
  Five-Sphere}''},
\doiref{10.1103/PhysRevLett.113.141601}{Phys.~Rev.~Lett. \textbf{113}, 141601
  (2014)\ignorespaces}\ignorespaces,
\normalsize{\texttt{\arxivref{1404.1925}{arXiv:1404.1925}}}\ignorespaces
\bibitem{Alday:2014bta}
L.~F. Alday, M.~Fluder, C.~M. Gregory, P.~Richmond \& J.~Sparks,
\textit{``{Supersymmetric gauge theories on squashed five-spheres and their
  gravity duals}''},
\doiref{10.1007/JHEP09(2014)067}{JHEP \textbf{1409}, 067
  (2014)\ignorespaces}\ignorespaces,
\normalsize{\texttt{\arxivref{1405.7194}{arXiv:1405.7194}}}\ignorespaces
\bibitem{Alday:2015jsa}
L.~F. Alday, M.~Fluder, C.~M. Gregory, P.~Richmond \& J.~Sparks,
\textit{``{Supersymmetric solutions to Euclidean Romans supergravity}''},
\doiref{10.1007/JHEP02(2016)100}{JHEP \textbf{1602}, 100
  (2016)\ignorespaces}\ignorespaces,
\normalsize{\texttt{\arxivref{1505.04641}{arXiv:1505.04641}}}\ignorespaces
\bibitem{Bern:2012di}
Z.~Bern, J.~J. Carrasco, L.~J. Dixon, M.~R. Douglas, M.~von~Hippel \&
  H.~Johansson,
\textit{``{D=5 maximally supersymmetric Yang-Mills theory diverges at six
  loops}''},
\doiref{10.1103/PhysRevD.87.025018}{Phys.~Rev. \textbf{D87}, 025018
  (2013)\ignorespaces}\ignorespaces,
\normalsize{\texttt{\arxivref{1210.7709}{arXiv:1210.7709}}}\ignorespaces
\bibitem{Bossard:2015foa}
G.~Bossard \& A.~Kleinschmidt,
\textit{``{Loops in exceptional field theory}''},
\doiref{10.1007/JHEP01(2016)164}{JHEP \textbf{1601}, 164
  (2016)\ignorespaces}\ignorespaces,
\normalsize{\texttt{\arxivref{1510.07859}{arXiv:1510.07859}}}\ignorespaces
\bibitem{Bossard:2017kfv}
G.~Bossard \& A.~Kleinschmidt,
\textit{``{Cancellation of divergences up to three loops in exceptional field
  theory}''},
\doiref{10.1007/JHEP03(2018)100}{JHEP \textbf{1803}, 100
  (2018)\ignorespaces}\ignorespaces,
\normalsize{\texttt{\arxivref{1712.02793}{arXiv:1712.02793}}}\ignorespaces
\bibitem{Bossard:2010pk}
G.~Bossard, P.~S. Howe, U.~Lindstrom, K.~S. Stelle \& L.~Wulff,
\textit{``{Integral invariants in maximally supersymmetric Yang-Mills
  theories}''},
\doiref{10.1007/JHEP05(2011)021}{JHEP \textbf{1105}, 021
  (2011)\ignorespaces}\ignorespaces,
\normalsize{\texttt{\arxivref{1012.3142}{arXiv:1012.3142}}}\ignorespaces
\bibitem{Tizzano:2014roa}
L.~Tizzano \& J.~Winding,
\textit{``{Multiple sine, multiple elliptic gamma functions and rational
  cones}''},
\normalsize{\texttt{\arxivref{1502.05996}{arXiv:1502.05996}}}\ignorespaces
\end{thebibliography}

\end{document}